\documentclass[twocolumn]{aastex631}
\usepackage{amsmath}
\usepackage{multirow}

\newcommand{\chandra}{{\it Chandra}}
\newcommand{\xmm}{{\it XMM-Newton}}
\newcommand{\hst}{{\it HST}}
\newcommand{\spitzer}{{\it Spitzer}}
\newcommand{\swift}{{\it Swift}}
\newcommand{\prospector}{{\sc Prospector}}
\newcommand{\galex}{{\it GALEX}}

\newcommand{\Msun}{M$_{\odot}$}

\newcommand{\Lx}{$L_{\rm X}$}

\newcommand{\flux}{erg s$^{-1}$ cm$^{-2}$}
\newcommand{\lum}{erg s$^{-1}$}

\newcommand{\Mstar}{M$_{\star}$}

\shortauthors{Binder et al.}

\begin{document}

\title{A Panchromatic Study of the X-ray Binary Population in NGC~300 on Sub-Galactic Scales}

\correspondingauthor{Breanna A. Binder}
\email{babinder@cpp.edu}

\author[0000-0002-4955-0471]{Breanna A. Binder}
\affiliation{Department of Physics \& Astronomy, California State Polytechnic University, 3801 W. Temple Ave, Pomona, CA 91768, USA}

\author{Rosalie Williams}
\affiliation{Department of Physics \& Astronomy, California State Polytechnic University, 3801 W. Temple Ave, Pomona, CA 91768, USA}

\author{Jacob Payne}
\affiliation{Department of Physics \& Astronomy, California State Polytechnic University, 3801 W. Temple Ave, Pomona, CA 91768, USA}

\author[0000-0002-3719-940X]{Michael Eracleous}
\affiliation{Department of Astronomy \& Astrophysics and Institute for Gravitation and the Cosmos, The Pennsylvania State University, 525 Davey Lab, University Park, PA 16802, USA}

\author[0000-0002-8606-0797]{Alexander Belles}
\affiliation{Department of Astronomy \& Astrophysics and Institute for Gravitation and the Cosmos, The Pennsylvania State University, 525 Davey Lab, University Park, PA 16802, USA}

\author[0000-0002-7502-0597]{Benjamin F. Williams}
\affiliation{Department of Astronomy, University of Washington, Box 351580, Seattle, WA 98195, USA}

\begin{abstract}
The population-wide properties and demographics of extragalactic X-ray binaries (XRBs) correlate with the star formation rates (SFRs), stellar masses (\Mstar), and environmental factors (such as metallicity, $Z$) of their host galaxy. Although there is evidence that XRB scaling relations (\Lx/SFR for high mass XRBs [HMXBs] and \Lx/\Mstar\ for low mass XRBs [LMXBs]) may depend on metallicity and stellar age across large samples of XRB-hosting galaxies, disentangling the effects of metallicity and stellar age from stochastic effects, particularly on subgalactic scales, remains a challenge. We use archival X-ray through IR observations of the nearby galaxy NGC~300 to self-consistently model the broadband spectral energy distribution and examine radial trends in its XRB population. We measure a current ($<$100 Myr) SFR of 0.18$\pm$0.08 \Msun\ yr$^{-1}$ and \Mstar= $(2.15^{+0.26}_{-0.14})\times10^9$ \Msun. Although we measure a metallicity gradient and radially resolved star formation histories that are consistent with the literature, there is a clear excess in the number of X-ray sources below $\sim10^{37}$ \lum\ that are likely a mix of variable XRBs and additional background AGN. When we compare the subgalactic \Lx/SFR ratios as a function of $Z$ to the galaxy-integrated \Lx-SFR-$Z$ relationships from the literature, we find that only the regions hosting the youngest ($\lesssim$30 Myr) HMXBs agree with predictions, hinting at time evolution of the \Lx-SFR-$Z$ relationship.
\end{abstract}

\keywords{X-ray binary stars (1811) --- High-mass X-ray binary stars (733) --- Compact objects (288) --- Spiral galaxies (1560)}

\section{Introduction} \label{sec:intro}
Some of the most useful laboratories for studying binary star evolution are X-ray binaries (XRBs), which contain a compact object (either a neutron star, NS, or a black hole, BH) accreting mass from a stellar companion. High-mass XRBs (HMXBs) contain a massive ($\geq8$\Msun), often post-main sequence, companion and represent a critical phase in the history of massive star evolution: their current properties are the result of the orbital and mass-exchange history of the progenitor binary, and the ultimate fate of some of these binaries is to form millisecond pulsars, short gamma-ray bursts (GRBs), or double compact object systems capable of generating detectable gravitational wave (GW) signals. Feedback from HMXBs is now recognized as an important contributor to heating the intergalactic medium and re-ionizing the early universe \citep{Kaur+22}. 

It is now well-established that the X-ray luminosity functions (XLFs) and population-integrated luminosities (\Lx) of HMXBs scale with host galaxy star formation rate \citep[SFR;][]{Grimm+03,Ranalli+03,Gilfanov+04,Mineo+12}, while those of low-mass XRBs (LMXBs, which typically contain companions of $\sim$1-2 \Msun) track host galaxy stellar mass \citep[\Mstar;][]{Boroson+11,Zhang+12, Lehmer+10,Gilfanov04,Colbert+04}. Previous studies have attempted to empirically measure the dependence of XRB scaling relations ($L_{\rm X}^{\rm HMXB}$/SFR or $L_{\rm X}^{\rm LMXB}$/\Mstar) on physical parameters such as age \citep{Kim+10,Lehmer+14} and metallicity \citep{BasuZych+13,BasuZych+16,Brorby+16,Tzanavaris+16, Douna+15}. Population synthesis models predict that the these scaling relations in XRB populations can vary by up to an order of magnitude depending on the metallicity and age of the XRB population \citep{Linden+10,Fragos+08,Fragos+13b,Fragos+13a,Zuo+14}. Nearby galaxies, where XRB populations can be spatially resolved with X-ray telescopes such as \chandra\ and \xmm, span a relatively narrow range in metallicities while hosting a larger diversity of stellar ages when compared to distant galaxies, and have experienced a wide variety of star formation histories (SFHs) and merger/interaction histories. As such, we do not yet precisely understand how the XRB scaling relations vary with SFH and metallicity.

Studies of the HMXB \Lx/SFR fraction as a function of metallicity \citep[hereafter referred to as the \Lx-SFR-$Z$ relation; e.g.,][]{Brorby+14,BasuZych+16} have largely concentrated on high-\Mstar\ galaxies with high-SFRs where the XRB populations are expected to be dominated by HMXBs and the XLFs are well-sampled. However, these studies implicitly assume that the {\it global} average SFR and metallicity of galaxies adequately reflect the sub-kpc scale environments in which HMXBs form. Due to their short lifetimes, HMXBs do not migrate more than a few hundred parsecs from their birthplaces \citep{Bodaghee+12,Binder+23} and these temporal and spatial scales are shorter than those over which chemical mixing efficiently occurs in galaxies \citep{Kreckel+20}. Local variations in stellar populations and the HMXBs associated with them thus become ``smeared out'' on galaxy-wide scales, and may complicate interpretations of the \Lx/SFR scaling relation \citep{Kouroumpatzakis+20}. Optical integral field units now make it possible to study thousands of sub-kiloparsec scale regions across the star-forming disks of nearby galaxies, and 2D metallicity maps demonstrate that metallicity variations occur on small scales \citep{Williams+22}, and many nearby galaxies exhibit metallicity gradients \citep[see, e.g.,][and references therein]{Belfiore+19}. 

The nearby \citep[2.0 Mpc, based on measurements of the tip of the red giant branch;][]{Dalcanton+09} star-forming galaxy NGC~300 is a particularly suitable target for studying XRBs in subgalactic environments. Numerous X-ray point sources are detected within the star-forming disk, which can be imaged in a single \chandra\ pointing down to limiting luminosities on the order of $\sim$10$^{36}$ \lum\ \citep{Binder+12,Binder+17}, it has been observed across nearly the entire electromagnetic spectrum, and exhibits a metallicity gradient with metallicity decreasing by $\sim$0.08$Z_{\odot}$ per kpc;][]{Gogarten+10,Kudritzki+08}. The SFR of NGC~300 is generally reported to be $\sim0.08-0.46$ \Msun\ yr$^{-1}$ by different tracers, such as the \hst\ resolved stellar populations \citep{Gogarten+10}, UV observations \citep{Mondal+19}, H$\alpha$ emission \citep{Karachentsev+13}, and mid-IR observations \citep{Helou+04}. NGC~300 contains a ULX \citep[e.g., see review by][and references therein]{Binder+20} and the luminous Wolf-Rayet + BH binary NGC~300 X-1 \citep{Binder+21}, although the majority of XRBs in the galaxy are of lower luminosity \citep[$\sim10^{36-37}$ \lum;][]{Binder+12,Binder+17}, and the specific SFR of NGC~300 suggests that its XRB population is likely a mix of HMXBs and LMXBs \citep{Lehmer+19}.

We use archival multiwavelength observations to self-consistently model the stellar mass (\Mstar), SFH, metallicity, and XRB population of NGC~300 on subgalactic scales and search for trends in the \Lx/SFR ratio for HMXBs as a function of metallicity and stellar age. The paper is organized as follows: in Section~\ref{sec:SEDs} we summarize the archival FUV-IR observations used and our methodology for performing aperture photometry, and present spectral energy distributions (SEDs) for NGC~300. We additionally describe how the SED modeling package \prospector\ \citep{Leja+17} is used to extract SFHs, stellar masses, and metallicites for the subgalactic regions within NGC~300. In Section~\ref{sec:data_xray} we describe the \chandra\ X-ray observations and compare the observed X-ray point source population to predicted populations based on prior studies of the XLFs of HMXBs, LMXBs, and the AGN log$N$-log$S$ distribution. We discuss our results in Section~\ref{sec:discussion}, and conclude with a summary of our main findings in Section~\ref{sec:conclusion}. All uncertainties correspond to the 90\% confidence interval unless otherwise specified. Table~\ref{tab:NGC300} summarizes the main properties of NGC~300 that we assume throughout this work, including the distance, central right ascension and declination coordinates, Galactic absorbing column, $A_V$, and position and inclination angles. The distance to NGC~300 was measured directly by \citet{Dalcanton+09} using the tip of the red giant branch, and is thus independent of cosmology.

\begin{deluxetable*}{cccccccc}
    \caption{Basic Properties of NGC~300}\label{tab:NGC300}
    \tablehead{
    \colhead{Distance\tablenotemark{a}} & \colhead{R.A.} & \colhead{Dec.} & \colhead{$N_{\rm H, Gal}$\tablenotemark{b}} & \colhead{$A_V$\tablenotemark{c}} & \colhead{$E(B-V)$\tablenotemark{c}} & \colhead{Position} & \colhead{Inclination} \\ \cline{2-3}
    (Mpc)   &  \multicolumn{2}{c}{(J2000)}     & (cm$^{-2}$)     & (mag)    & (mag)     & Angle\tablenotemark{d} ($^{\circ}$)     & Angle\tablenotemark{d} ($^{\circ}$)
    }
    \colnumbers
    \startdata
    2.0 & 00:54:53.48  & -37:41:03.8        & 10$^{21}$ & 0.035   & 0.011 & 111   & 45 \\
    \enddata
    \tablerefs{$^a$\citet{Dalcanton+09}, $^b$\citet{HI4PICollaboration+16}, $^c$\citet{Schlafly+11}, $^d$\citet{RC3}}
\end{deluxetable*}

\vspace{-0.5cm}
\section{Spectral Energy Distributions} \label{sec:SEDs}
\subsection{The Data}
In this section we summarize the publicly-available multiwavelength observations used to build the SEDs of NGC~300. Foreground stars were identified using the Gaia DR3 catalog \citep{GaiaDR3} and through visual inspection of multiwavelength images and then masked with circular apertures of radii of $\sim$15$^{\prime\prime}$. The median flux density per unit area of the corresponding subgalactic region in which each star was located was used to approximate the the flux density of the galaxy in the masked region.

We note that, although NGC~300 has been imaged many times with the \hst, the archival \hst\ observations do not optimally cover the star-forming disk of NGC~300 and were obtained for a variety of filter configurations. For example, \hst/F606W imaging covers only $\sim$50\% of the star-forming disk of NGC~300, while F475W imaging covers less than $\sim$20\% of the disk. Inferring galaxy-integrated optical flux densities from \hst\ would require a significant degree of extrapolation and carry significant uncertainty, limiting their utility for SED modeling. We therefore do not include these data in our SED modeling.

\subsubsection{Spitzer IRAC and MIPS}\label{sec:data_ir}
We first retrieved \spitzer\ IRAC and MIPS mosaics from the NASA/IPAC Infrared Science Archive (IRSA)\footnote{\url{https://irsa.ipac.caltech.edu/data/SPITZER/Enhanced/SEIP/overview.html}}. The IRAC channels have central wavelengths of to 3.6$\mu$m, 4.5$\mu$m, 5.8$\mu$m, and 8.0$\mu$m, while the MIPS channels have central wavelengths of 24$\mu$m, 70$\mu$m, and 160$\mu$m. The IRAC images have an angular resolution $\sim2^{\prime\prime}$, while the angular resolution of the MIPS images are $\sim6^{\prime\prime}$, $\sim18^{\prime\prime}$, and $\sim40^{\prime\prime}$ at 24$\mu$m, 70$\mu$m, and 160$\mu$m, respectively \citep{Rieke+04}. All \spitzer\ images have units of MJy sr$^{-1}$.

\begin{figure}
    \centering
    \includegraphics[width=1\linewidth]{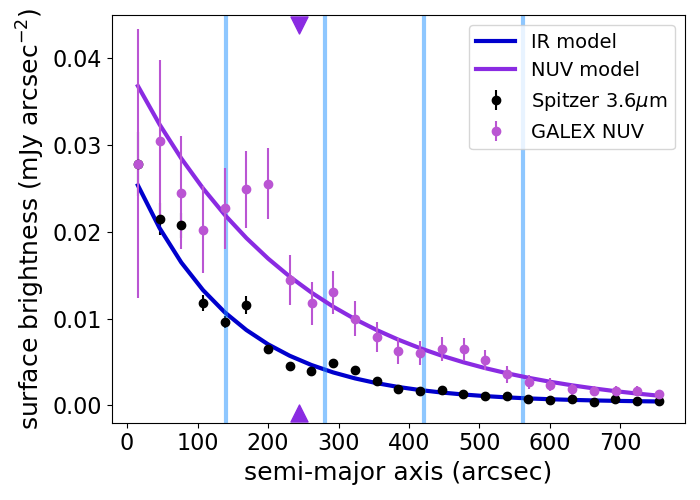}
    \caption{The \spitzer/IRAC 3.6$\mu$m surface brightness profile (in mJy arcsec$^{-2}$; data in black, best-fit exponential disk model in blue) of NGC~300, with the four radial bins (each with a width of $R_d=1.36$ kpc) indicated by vertical light blue lines. The GALEX NUV surface brightness (normalized to the peak value of the 3.6$\mu$m profile) is shown in purple. The purple upward and downward triangle indicates the scale length of the the NUV disk.}
    \label{fig:surfbri}
\end{figure}

We measured the surface brightness profile of NGC~300 using the the IRAC1 (3.6$\mu$m) image. We constructed a set of 25 elliptical annuli centered on the galaxy center and using a position angle of 111$^{\circ}$ for the major axis. The resulting surface brightness profile, in units of mJy arcsec$^{-2}$, is shown by the black dots in Figure~\ref{fig:surfbri}. We modeled the surface brightness profile with an fit an exponentially decaying function, 

\begin{equation}
    I(r) = I_0 e^{-r/R_d} + I_{\rm bkg},
\end{equation}

\noindent using SciPy's \texttt{curve\_fit} package, where $I_0$ is the central surface brightness, $R_d$ is the disk scale length, and $I_{\rm bkg}$ is the background level surface brightness. We find best-fit parameters $I_0=(2.79\pm0.14)\times10^{-5}$ Jy arcsec$^{-2}$, $R_d=(140.3\pm4.7)$ arcsec (corresponding to a physical length of $1.36\pm0.05$ kpc), and $I_{\rm bkg}=(3.20\pm0.47)\times10^{-7}$ Jy arcsec$^{-2}$ (see the blue line in Figure~\ref{fig:surfbri}). This scale length is consistent with those measured in previous optical and IR studies \citep{Gogarten+10,Jang+20}. The UV disk of NGC~300 is known to be more extended than the optical/IR disk \citep[$\sim$2.7 kpc in the NUV, see][]{Mondal+19}. For comparison, we also show the GALEX NUV surface brightness profile in Figure~\ref{fig:surfbri}, for which we measure a scale length of $\sim$2.4 kpc (see Section~\ref{sec:data_uv} for a discussion of GALEX data.)

The 3.6$\mu$m scale length was then used to divide the galaxy into four equal-width radial bins. We opted for the 3.6$\mu$m scale length, rather than the more extended NUV scale length for three main reasons: (1) the smaller annular regions reduce the amount of ``smearing out'' of key environmental factors such at metallicity and star formation history, both of which are known to vary across the NGC~300 disk \citep{Kudritzki+08,Gogarten+10}; (2) since the expected AGN contribution increases with surface area, utilizing smaller annular regions reduces the expected contribution from AGN within each ring; and (3) at many wavelengths the relative uncertainty in the fainter flux densities at large galactocentric radius results in more uncertain SED models (see Section~\ref{sec:SEDs}). The four radial bins are an ``inner'' bin ($r\leq R_d$), an ``inner middle'' bin ($R_d < r \leq 2R_d$), an ``outer middle'' bin ($2R_d < r \leq 3R_d$), and an ``outer'' bin ($3R_d < r \leq 4R_d$). At a distance of $4R_d$, the surface brightness falls to $e^{-4}\approx1.8$\% of the central value; the uncertainties in the total flux density beyond $4R_d$ are large, diminishing our ability to confidently constrain the metallicity, stellar mass, and SFR in these outer regions.

\subsubsection{WISE}\label{sec:data_wise}
We used the WISE Image Service hosted on the NASA/IPAC Infrared Science Archive\footnote{\url{https://irsa.ipac.caltech.edu/Missions/wise.html}} to retrieve WISE images in bands W1 (3.4 $\mu$m), W2 (4.6 $\mu$m), W3 (12 $\mu$m) and W4 (22$\mu$m). The angular resolution is 6.1$^{\prime\prime}$, 6.4$^{\prime\prime}$, 6.5$^{\prime\prime}$ and 12$^{\prime\prime}$ for the W1, W2, W3, and W4 bands, respectively \citep{Wright+10}. Instrumental magnitudes are converted to AB magnitudes using the magnitude zero-points for each filter (in order of increasing central wavelength: 20.5 mag, 19.5 mag, 18.0 mag, and 13.0 mag) and the AB magnitude offset defined for each band \citep[in order of increasing central wavelength: 2.70 mag, 3.34 mag, 5.17 mag, 6.62 mag;][]{Cutri+13}.

\subsubsection{2MASS}\label{sec:data_2mass}
We retrieved background-subtracted FITS images of NGC~300 from the 2MASS Large Galaxy Atlas\footnote{\url{https://irsa.ipac.caltech.edu/applications/2MASS/LGA/intro.html}} \citep{Jarrett+03}. The $J$, $H$, and $K_s$ bands used by 2MASS are particularly sensitive to older stellar populations, which dominate the stellar mass of NGC~300 \citep{Gogarten+10}. The 2MASS images have an angular resolution of $\sim4^{\prime\prime}$ and the photometric zero-point calibration of the images is accurate to $\sim2-3$\%.

\vspace{-0.01cm}
\subsubsection{Swift Ultraviolet/Optical Telescope}\label{sec:data_swift}
We retrieved all available data from the Ultraviolet/Optical Telescope on the Neil Gehrels \swift\ Observatory from the High Energy Astrophysics Science Archive Research Center (HEASARC)\footnote{\url{https://heasarc.gsfc.nasa.gov/cgi-bin/W3Browse/swift.pl}} that are within 17 arcminutes of NGC 300 \citep{2004ApJ...611.1005G, 2005SSRv..120...95R}. For each of the six filters (UVW2, UVM2, UVW1, u, b, v), we processed the data by correcting for large scale sensitivity variations and masking bad pixels. Prior to stacking the data, we corrected each individual exposure for the time dependent detector sensitivity loss \footnote{\url{https://heasarc.gsfc.nasa.gov/docs/heasarc/caldb/swift/docs/uvot/uvotcaldb_throughput_06.pdf}} and subtracted off the necessary counts to normalize the background count rate across exposures. Normalizing the background is especially important in the optical filters due to effects of scattered light from the sun, Earth, and moon. In total, the final stacked images included 151, 129, 136, 94, 24.3, 18.5 ks for the UVW2, UVM2, UVW1, u, b, and v filters respectively. The angular resolution of the UVOT filters is $\sim2.5^{\prime\prime}$.

As \swift\ was designed to study point sources, photometry must be done with care. Using the apertures defined in Section~\ref{sec:data_ir}, we calculated the accumulated count rate as a function of radius, following \citet{2007ApJS..173..185G}. This brightness profile was then used to determine the magnitude in each radial bin using the zero-point offsets from \citet{2011AIPC.1358..373B}. As discussed in \citet{Decleir+19} and \citet{2023arXiv230505650B}, the zero-point offsets are defined based on a 5 arcsecond aperture. Therefore, we divided our measured fluxes by the asymptote of the curve of growth \citep{2010MNRAS.406.1687B} in order to account for the difference in the effective aperture area. We used the Python package \texttt{dust\_extinction}\footnote{\url{https://dust-extinction.readthedocs.io/en/stable/}} and the ``F99'' dust model from \citet{Fitzpatrick99}, along with $A_V$ and $E(B-V)$ provided in Table~\ref{tab:NGC300}, to correct the \swift\ flux densities for dust extinction and attenuation. 

\vspace{-0.1cm}
\subsubsection{GALEX}\label{sec:data_uv}
We retrieved \galex\ FUV and NUV imaging from the Mikulski Archive for Space Telescopes (MAST:  \dataset[10.17909/k7y9-yp18]{http://dx.doi.org/10.17909/k7y9-yp18}). The images were obtained on 2004 October 26, with a total exposure time of 12987.6 s in both filters. The angular resolution is $4.3^{\prime\prime}$ in the FUV and $5.3^{\prime\prime}$ in the NUV, and images are in units of counts per second. Images are converted to flux density units of \flux\ \AA$^{-1}$ by multiplying by $1.4\times10^{-15}$ (FUV) or $2.06\times10^{-16}$ \citep[NUV; ][]{Martin+05, Morrissey+07}. As for the \swift/UVOT observations (Section~\ref{sec:data_swift}), the \galex\ flux densities are corrected for Galactic dust extinction and attenuation.

\subsection{Modeling the SED}
\begin{figure}
    \centering
    \includegraphics[width=1\linewidth,clip=true,trim=2cm 5.5cm 2cm 5cm]{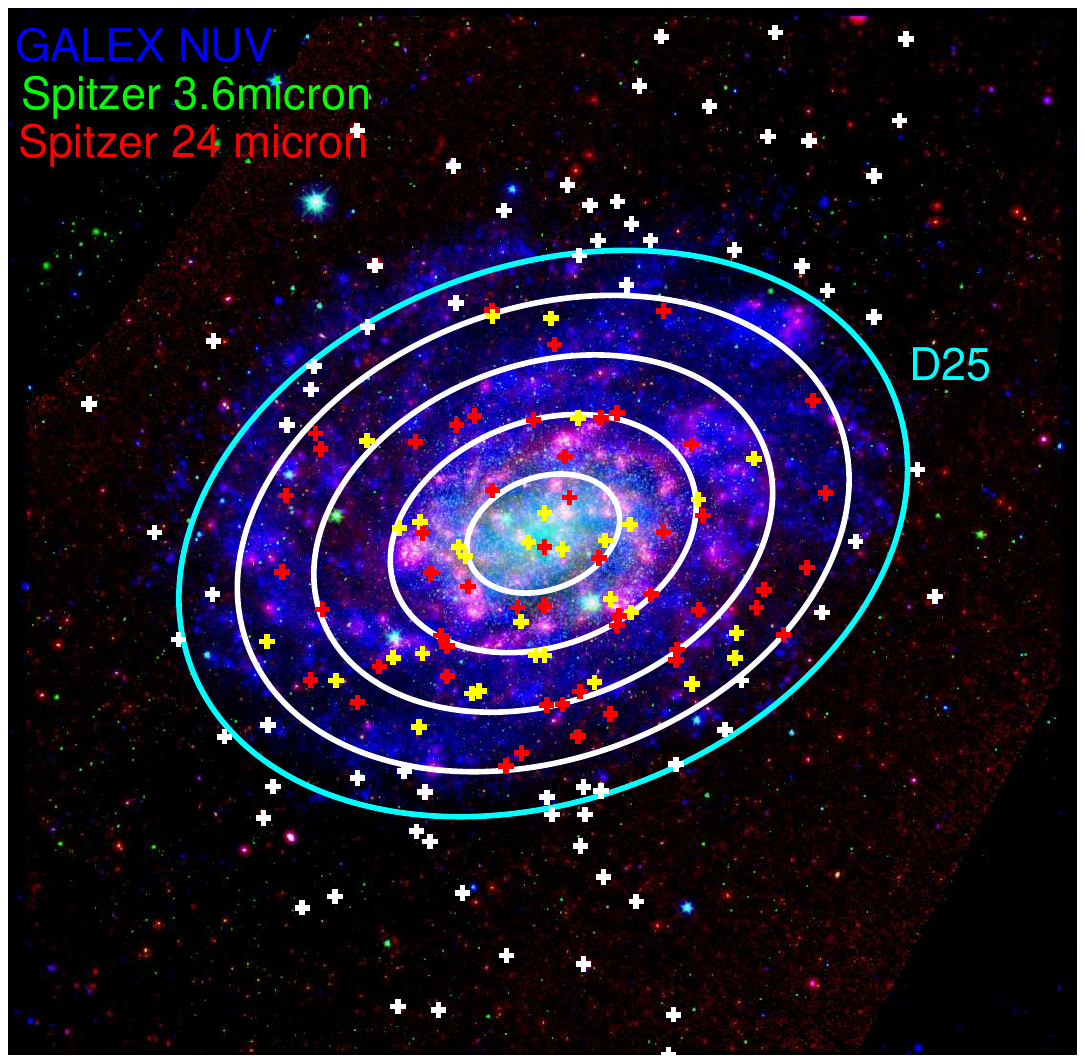}
    \caption{An RGB-rendered image of NGC~300, with the with four subgalactic regions shown by white elliptical annuli. The $D_{25}$ isophote \citep{RC3} is shown in cyan for comparison. The \galex\ NUV image is shown in the blue channel, the \spitzer\ IRAC 3.6$\mu$m is shown in green, and the \spitzer\ MIPS 24$\mu$m is shown in red. \chandra\ X-ray sources that heavily absorbed and likely to be AGN are shown inred, sources that are consistent with being XRBs are shown as yellow crosses, and X-ray sources that are not used are shown as white crosses (see Section~\ref{sec:data_xray}).}
    \label{fig:RGB}
\end{figure}

Figure~\ref{fig:RGB} shows an RGB-rendered image of NGC~300 (using \galex\ NUV, \spitzer\ 3.6$\mu$m, and \spitzer\ 24$\mu$m imaging in the blue, green, and red channels, respectively) with the elliptical annuli corresponding to a width of $R_d$ superimposed (the locations of \chandra\ X-ray sources used in our XRB analysis are shown in yellow, while X-ray sources that are not used are shown in white; see Section~\ref{sec:data_xray}). The $D_{25}$ isophotal ellipse \citep{RC3}, shown in cyan for comparison, is slightly larger than the outer annular region. Table~\ref{tab:photometry} provides the photometry measurements that are used to construct SEDs for the entire galaxy and the four subgalactic annuli. We use \prospector\ \citep{Leja+17,Johnson+21}\footnote{See also \url{https://github.com/bd-j/prospector}} to model the SEDs of NGC~300. \prospector\ forward-models the observed photometry given parameters that describe the gas, dust, and stellar populations present in the galaxy in addition to instrumental parameters. The code uses noise models and prior distributions to compute the likelihood and posterior probability for the given model and performs Monte Carlo sampling of the posterior probability distribution. \prospector\ has been extensively tested on SEDs and spectroscopy of nearby galaxies \citep{Leja+17,Leja+18}, dwarf galaxies \citep{Greco+18,Pandya+18}, and has been used in spatially-resolved studies \citep{Patricio+18,Oyarzun+19}.

We use the non-parametric ``continuity star formation history'' (\texttt{continuity\_sfh}) model \citep{Leja+19,Johnson+21} to fit piecewise constant (or step function) SFHs in seven time bins. Rather than directly fitting for the SFR in each time bin, this model utilizes parameters that describe the ratio of SFRs in adjacent temporal bins. The default priors on these ratios are described by Student-$t$ distributions and a \citet{Kroupa01} IMF. This model results in SFHs that tend towards constant SFRs, as dramatic differences in SFRs between adjacent temporal bins are down-weighted. We selected this SFH model given that NGC~300 shows no evidence of having undergone dramatic merger or interaction events that would lead to brief, high-SFR episodes. We adopt the same time bins as \citet{Leja+19}: 0-30 Myr, 30-100 Myr, 100-330 Myr, 330 Myr-1.1 Gyr, 1.1-3.6 Gyr, 3.6-11.7 Gyr, and 11.7-13.7 Gyr. The first two bins are directly relevant to the current HMXB population, while the last time bin is selected to permit a maximally old population. The intermediate time bins spanning from 100 Myr to 11.7 Gyr are equally spaced in logarithmic time \citep[see, e.g.,][]{Ocvirk+06}.

\vspace*{-\baselineskip}\vspace*{-\baselineskip}
\begin{deluxetable*}{ccccccc}
    \caption{UV, Optical, and IR Photometry of NGC~300}\label{tab:photometry}
    \tablehead{
                        &                       & \multicolumn{5}{c}{Flux Densities (Jy)} \\ \cline{3-7}
    \colhead{Instrument} & \colhead{Wavelength} & \colhead{Whole Galaxy}  & \colhead{Inner} & \colhead{Inner Middle} & \colhead{Outer Middle} & \colhead{Outer}}
    \colnumbers
    \startdata
    \galex\ FUV & 1535.08 \AA    & 0.231$\pm$0.003 & 0.026$\pm$0.001 & 0.084$\pm$0.001 & 0.067$\pm$0.002 & 0.054$\pm$0.003 \\
    \swift\ UVW2 & 2054.61 \AA   & 0.261$\pm$0.005 & 0.035$\pm$0.001 & 0.096$\pm$0.003 & 0.076$\pm$0.003 & 0.056$\pm$0.002 \\
    \swift\ UVM2 & 2246.43 \AA   & 0.253$\pm$0.005 & 0.034$\pm$0.001 & 0.093$\pm$0.003 & 0.074$\pm$0.003 & 0.054$\pm$0.002 \\
    \galex\ NUV & 2300.78 \AA    & 0.316$\pm$0.003 & 0.039$\pm$0.001 & 0.111$\pm$0.001 & 0.091$\pm$0.003 & 0.074$\pm$0.003 \\
    \swift\ UVW1 & 2580.74 \AA   & 0.305$\pm$0.007 & 0.047$\pm$0.002 & 0.113$\pm$0.004 & 0.086$\pm$0.004 & 0.059$\pm$0.004 \\
    \swift\ u   & 3467.05 \AA    & 0.621$\pm$0.007 & 0.099$\pm$0.002 & 0.247$\pm$0.005 & 0.158$\pm$0.004 & 0.118$\pm$0.003 \\
    \swift\ b   & 4349.56 \AA    & 1.288$\pm$0.028 & 0.247$\pm$0.006 & 0.421$\pm$0.013 & 0.351$\pm$0.017 & 0.269$\pm$0.018 \\
    \swift\ v   & 5425.33 \AA    & 1.945$\pm$0.018 & 0.421$\pm$0.005 & 0.635$\pm$0.009 & 0.508$\pm$0.010 & 0.382$\pm$0.011 \\
    2MASS $J$   & 1.235 $\mu$m   & 4.109$\pm$0.069 & 1.065$\pm$0.018 & 1.660$\pm$0.018 & 1.024$\pm$0.062 & 0.360$\pm$0.068 \\
    2MASS $H$   & 1.662 $\mu$m   & 4.047$\pm$0.064 & 1.076$\pm$0.019 & 1.500$\pm$0.019 & 1.018$\pm$0.057 & 0.453$\pm$0.063 \\
    2MASS $K_s$ & 2.159 $\mu$m   & 3.088$\pm$0.054 & 0.811$\pm$0.014 & 1.208$\pm$0.014 & 0.742$\pm$0.050 & 0.327$\pm$0.053 \\
    WISE W1     & 3.4 $\mu$m     & 1.948$\pm$0.004 & 0.458$\pm$0.001 & 0.723$\pm$0.003 & 0.494$\pm$0.004 & 0.273$\pm$0.005 \\
    \spitzer\ IRAC1 & 3.6 $\mu$m & 1.857$\pm$0.101 & 0.410$\pm$0.016 & 0.750$\pm$0.038 & 0.454$\pm$0.090 & 0.243$\pm$0.109 \\
    \spitzer\ IRAC2 & 4.5 $\mu$m & 1.251$\pm$0.049 & 0.285$\pm$0.013 & 0.480$\pm$0.045 & 0.304$\pm$0.073 & 0.181$\pm$0.086  \\
    WISE W2     & 4.6 $\mu$m     & 1.143$\pm$0.004 & 0.266$\pm$0.001 & 0.453$\pm$0.004 & 0.278$\pm$0.006 & 0.151$\pm$0.006 \\
    \spitzer\ IRAC3 & 5.8 $\mu$m & 1.115$\pm$0.464 & 0.271$\pm$0.121 & 0.507$\pm$0.248 & 0.247$\pm$0.109 & $<$1.299 \\
    \spitzer\ IRAC4 & 8.0 $\mu$m & 2.156$\pm$0.034 & 0.500$\pm$0.012 & 0.999$\pm$0.023 & 0.453$\pm$0.028 & 0.203$\pm$0.031 \\
    WISE W3     & 12 $\mu$m      & 1.949$\pm$0.097 & 0.434$\pm$0.025 & 0.902$\pm$0.057 & 0.419$\pm$0.068 & 0.195$\pm$0.080 \\
    WISE W4     & 22 $\mu$m      & 2.017$\pm$0.765 & 0.465$\pm$0.192 & 1.280$\pm$0.413 & 0.510$\pm$0.467 & 0.222$\pm$0.133 \\
    \spitzer\ MIPS1 & 24 $\mu$m  & 1.993$\pm$0.479 & 0.389$\pm$0.128 & 1.092$\pm$0.330 & 0.348$\pm$0.186 & $<$0.423 \\
    \spitzer\ MIPS2 & 70 $\mu$m  & 39.242$\pm$2.878 & 7.333$\pm$0.980 & 18.728$\pm$0.134 & 8.680$\pm$0.152 & 4.502$\pm$0.167  \\
    \spitzer\ MIPS3 & 160 $\mu$m & 144.855$\pm$6.201 & 24.264$\pm$2.997 & 57.777$\pm$0.123 & 39.179$\pm$0.139 & 23.635$\pm$0.148
    \enddata
\end{deluxetable*}

The metallicity, stellar mass formed, and dust attenuation optical depth ($\tau_{\rm dust}$) are additional free parameters in our model. Metallicity priors are assumed to be Gaussian. Although the slopes of the metallicity gradients reported by \citet{Kudritzki+08} and \citet{Gogarten+10} agree, their central metallicity values differ by $\sim$0.3$Z_{\odot}$ (see Table~\ref{tab:params_lit}). The central metallicity value depends on the calibration used, and the value reported by \citet{Kudritzki+08} was derived from a combination of abundance estimates from mostly Fe, Ti, and Cr spectral lines. The CMD fitting technique utilized by \citet{Gogarten+10} relies on synthetic photometry of theoretical stellar isochrones to determine the best metallicity for a given stellar population. Given the consistency of the slopes, \citet{Gogarten+10} adopt the \citet{Kudritzki+08} central value when discussing the implications of the metallicity gradient. We initialized the metallicity prior by setting the mean value equal to the direct measurement value derived from \citet{Kudritzki+08} at each ring midpoint. The uncertainty in the metallicity is computed from the uncertainty in the observed metallicity gradient and the effect of the finite thickness of the radial bin added in quadrature; we set the standard deviation of the Gaussian prior to twice this metallicity uncertainty value (typically $\approx$0.3). This standard deviation was large enough to be inclusive of the \citet{Gogarten+10} values across the disk. We note that we obtained similar metallicity values when we used an uninformative prior on the metallicity, but we opted to retain the Gaussian prior on the metallicity as it decreased the amount of time required by \prospector\ to obtain a good fit. 

NGC~300 is estimated to have a total stellar mass of $\sim2.4\times10^{9}$ \Msun\ \citep{Puche+90,Sheth+10,Querejeta+15}. We assume this mass is distributed in the same manner as the 3.6$\mu$m surface brightness profile when providing an initial estimate of the stellar mass contained in each radial bin, and we assume a log uniform prior on the stellar mass (with minimum and maximum stellar mass values set to $10^6$ \Msun\ and $10^{10}$ \Msun, respectively). The dust attenuation parameter assumes a flat (top-hat) prior with minimum and maximum values of 0.0 and 2.0. We experimented with changing the maximum and minimum values used by the dust attenuation and stellar mass priors, and with changing the initial values of the stellar mass and metallicity. Altering these values had no significant effect on the resulting fits. Changing the initial value of the stellar mass by up to a factor of $\sim$5 similarly did not significantly affect our results. The best-fit model was determined via $\chi^2$ minimization, and parameter values around the best-fit were sampled with an ensemble MCMC algorithm \citep{Goodman+10} using the \texttt{emcee} Python package \citep{Foreman-Mackey+13}. We initialize \texttt{emcee} with 128 walkers and 512 iterations to build up a distribution of SED models. The best-fit parameters and uncertainties were defined by the 16th, 50th, and 84th percentile quantities of each distribution. 

\begin{deluxetable}{ccc}
    \caption{Previous Spatially-Integrated Measurements of NGC~300 Physical Parameters}\label{tab:params_lit}
    \tablehead{
    \colhead{Method} & \colhead{Value} & \colhead{Reference} 
     }
    \colnumbers
    \startdata
    \multicolumn{3}{c}{Star Formation Rate (\Msun\ yr$^{-1}$)} \\
    \hline
    \prospector & 0.177 & this work \\
    FUV+NUV   & 0.46    & \citet{Mondal+19} \\
    FUV       & 0.32    & \citet{Lee+09} \\
    UV        & 0.28    & \citet{Kaisina+12} \\
    CMD fitting & 0.15  & \citet{Gogarten+10} \\
    H$\alpha$ &  0.14   & \citet{Helou+04} \\
    H$\alpha$ & 0.17    & \citet{Kaisina+12} \\
    NUV+70$\mu$m  & 0.09 & \citet{Mineo+12} \\
    8$\mu$m    & 0.08-0.11 & \citet{Helou+04} \\
    \hline \hline
    \multicolumn{3}{c}{Stellar Mass (\Msun)} \\
    \hline
    \prospector & 2.15$\times10^9$ & this work \\
    H~I    & 4.2$\times10^9$ & \citet{Puche+90} \\
    IR (S$^4$G) & 2.1$\times10^9$ & \citet{Querejeta+15} \\
    Optical+IR (S$^4$G)    & 2.0$\times10^9$       & \citet{Laine+16} \\
    \hline \hline
    \multicolumn{3}{c}{Central Metallicity ($Z/Z_{\odot}$) \tablenotemark{a}} \\
    \hline
    \prospector & 0.54 & this work \\
    12+log(O/H) & 0.76  & \citet{Bresolin+09} \\
    12+log(O/H) & 0.78  & \citet{Micheva+22} \\
    $[M/H]$ & 0.59 & \citet{Gogarten+10} \\
    $Z/Z_{\odot}$ & 0.87 & \citet{Kudritzki+08}
    \enddata
    \tablecomments{$^a$Metallicites have been converted to $Z/Z_{\odot}$ for ease of comparison.}
\end{deluxetable}

\subsection{Comparison to Previous Studies}
\begin{figure*}
    \centering
    \includegraphics[width=1\linewidth]{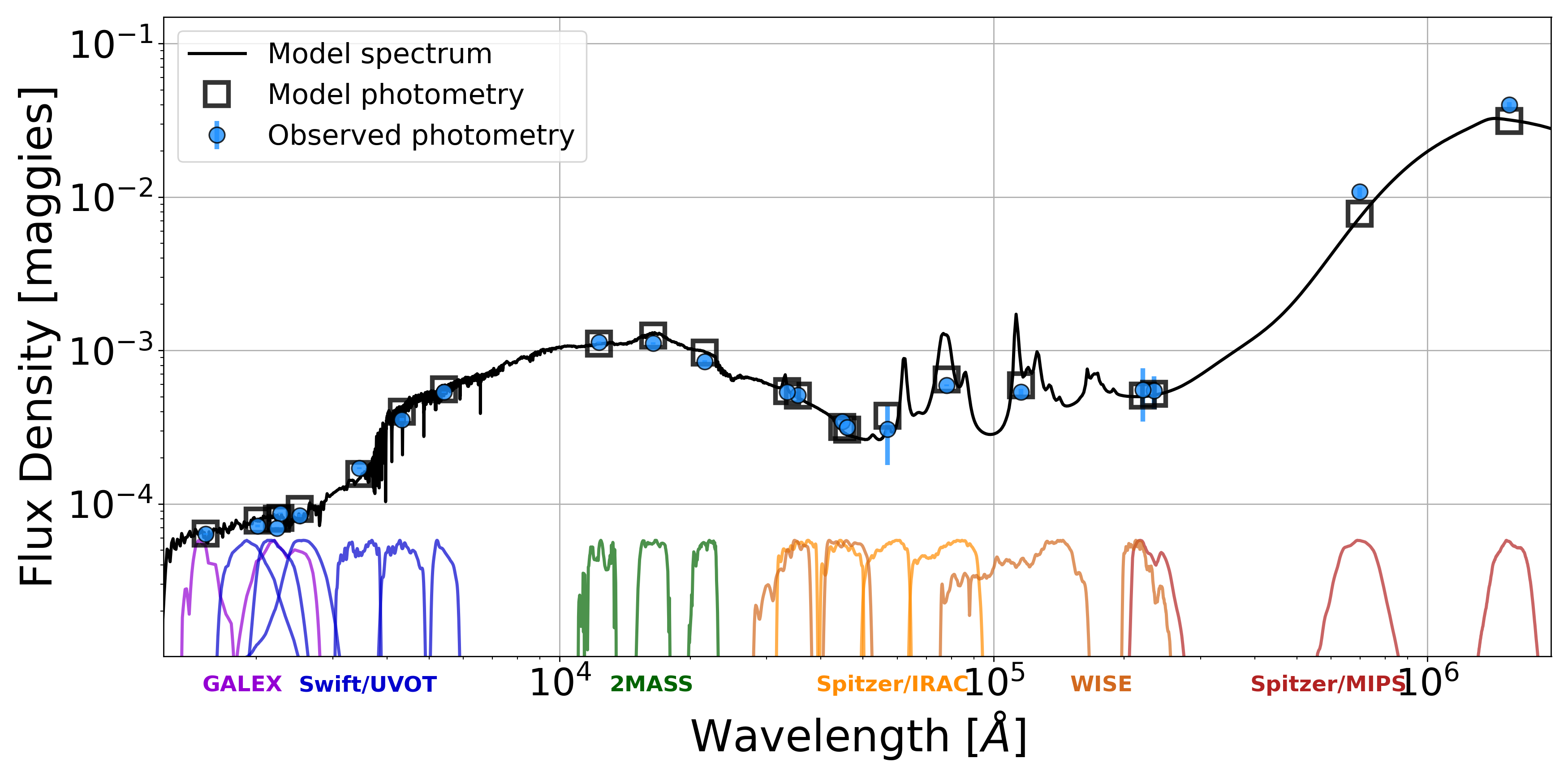}
    \caption{The best-fit, whole-galaxy (out to $4R_d$) SED for NGC~300. The observed photometry (from Table~\ref{tab:photometry}) is shown as blue dots. The best-fit model spectrum is shown in black, and black open squares indicate the model-predicted photometry for each filter used. The filter transmission curves are shown in multiple colors across the lower horizontal axis.}
    \label{fig:whole_SED}
\end{figure*}

\begin{figure}
    \centering
    \includegraphics[width=1\linewidth]{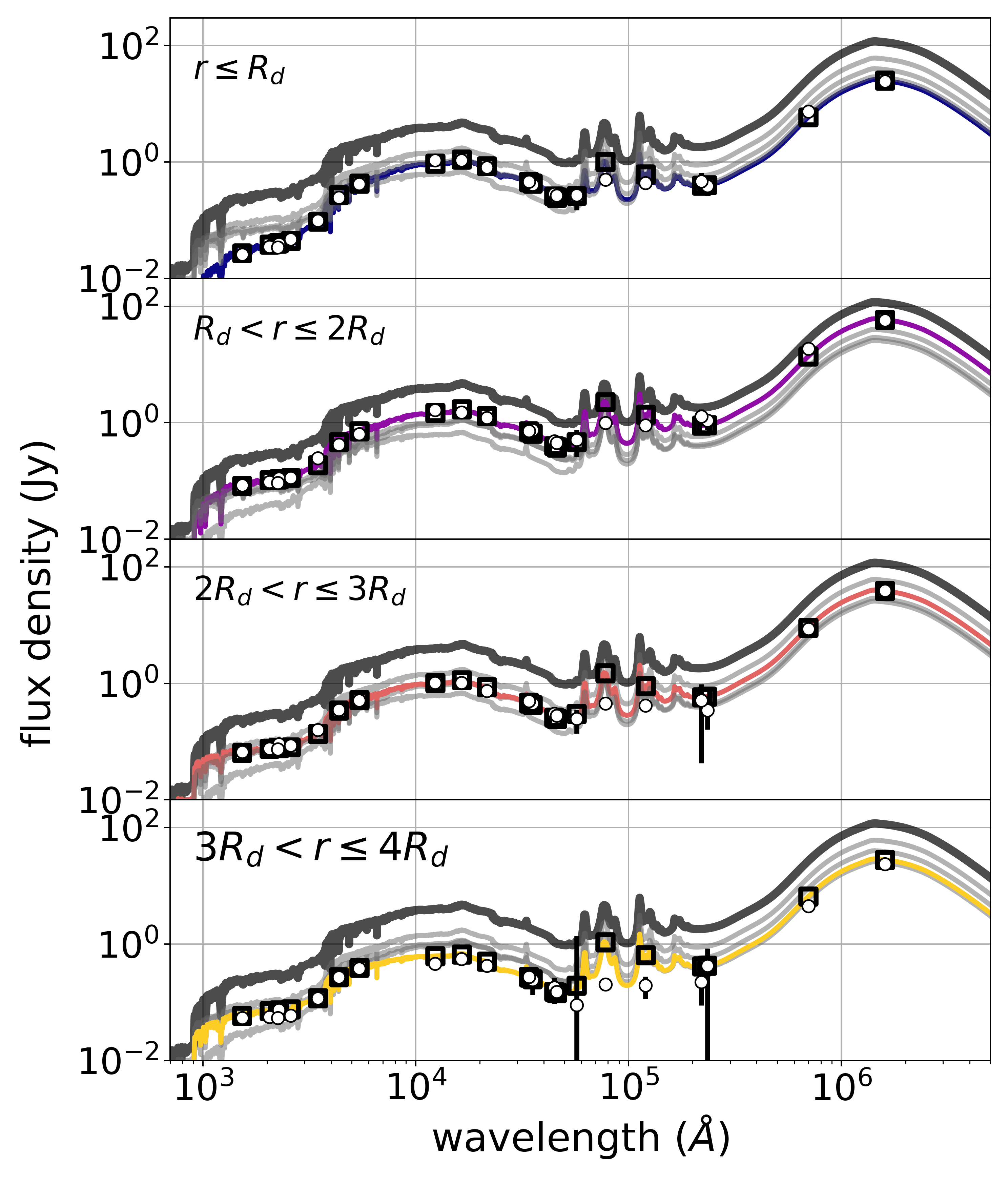}
    \caption{Best-fit radial SEDs from \prospector. All four SEDs are shown in each plot in gray, with the relevant sub-region highlighted in bold color. The galaxy-integrated best-fit SED model from Figure~\ref{fig:whole_SED} is shown by the thick, dark gray line. The white \textcolor{cyan}{{\bf circles}} show the photometry measurements from Table~\ref{tab:photometry}; in most cases, the uncertainties are smaller than the plotting symbol. Black squares show the predicted flux density in each filter.}
    \label{fig:SEDs}
\end{figure}

Table~\ref{tab:params_lit} provides an overview of the spatially-integrated SFR, $M_{\star}$, and central metallicity\footnote{\citet{Bresolin+09} and \citet{Micheva+22} report 12+log(O/H). We convert to $Z/Z_{\odot}$ assuming 10$^{X-8.69}$ \citep{Asplund+09}, where $X$ is the reported 12+log(O/H) value.} values reported for NGC~300 from the literature and our \prospector\ modeling; the detailed fit parameters and SFHs from \prospector\ are are summarized in Table~\ref{tab:SED_models}. We additionally provide the stellar mass formation history, which is the SFR in each time bin integrated over the duration of the time bin. This gives a measure of the total mass in stars that formed in each epoch; the current stellar mass \Mstar\ represents the surviving stellar mass at the present time and is lower than the total mass formed in stars. The best-fit SED parameters for the spatially-integrated SED and the radial subgalactic regions are modeled independently; i.e., column (3) of Table~\ref{tab:SED_models} is not the sum of columns (4)-(7). We note that the sums of the stellar masses and SFRs over the subgalactic regions yields values that are generally within $2\sigma$ of the value derived from the spatially-integrated SED. Figure~\ref{fig:whole_SED} shows the UV-to-IR galaxy-integrated SED of NGC~300 (out to 4$R_d$) along with the best-fit model from \prospector, while Figure~\ref{fig:SEDs} shows the best SED models and the corresponding UV-IR photometry for the four subgalactic radial bins. The best-fit model for the entire galaxy yields a stellar mass, recent ($<$100 Myr) SFR, and metallicity that are all consistent with values previously reported in the literature (and summarized in Table~\ref{tab:params_lit}).

\begin{deluxetable*}{cccccccc}
    \caption{Best-Fit SED Models from \prospector}\label{tab:SED_models}
    \tablehead{
     & & Galaxy Inte- & \multicolumn{4}{c}{Radial Bin} & Sum of All  \\ \cline{4-7}
     \colhead{Parameter}  & \colhead{Units} & \colhead{grated ($\leq$4$R_d$)} & \colhead{$\leq R_d$} & \colhead{$R_d<r \leq 2R_d$} & \colhead{$2R_d<r \leq 3R_d$} & \colhead{$3R_d<r \leq 4R_d$} & Radial Bins
     }
    \colnumbers
    \startdata
   \Mstar\         & 10$^9$ \Msun\     & 2.15$^{+0.26}_{-0.14}$    & 0.64$^{+0.03}_{-0.02}$    & 0.94$^{+0.08}_{-0.07}$ & 0.38$^{+0.06}_{-0.04}$ & 0.27$^{+0.03}_{-0.02}$ & 2.23$^{+0.20}_{-0.15}$ \\
    SFR($<$100 Myr) & \Msun\ yr$^{-1}$  & 0.177$\pm$0.070    & 0.016$^{+0.009}_{-0.007}$  & 0.094$^{+0.026}_{-0.033}$ & 0.045$^{+0.035}_{-0.026}$ & 0.014$^{+0.011}_{-0.006}$ & 0.169$^{+0.081}_{-0.072}$ \\
    log($Z$)        & log[$Z_{\odot}$]  & -0.239$^{+0.014}_{-0.033}$ & -0.266$^{+0.017}_{-0.013}$ & -0.360$\pm$0.013 & -0.459$^{+0.065}_{-0.046}$ & -0.628$^{+0.011}_{-0.012}$ & \nodata \\
    $\tau_{\rm dust}$     &         & 0.122$^{+0.008}_{-0.005}$ & 0.155$^{+0.007}_{-0.009}$ & 0.170$\pm$0.007 & 0.145$^{+0.009}_{-0.008}$ & 0.122$\pm$0.09 & \nodata  \\
    \hline \hline
    \multicolumn{7}{c}{Star Formation History}  \\
    \hline
    0-30 Myr    & \Msun\ yr$^{-1}$ & 0.048$\pm$0.018 & 0.005$^{+0.003}_{-0.002}$ & 0.016$^{+0.007}_{-0.005}$ & 0.017$^{+0.007}_{-0.006}$ & 0.013$^{+0.005}_{-0.004}$ & 0.051$^{+0.022}_{-0.017}$ \\
    30-100 Myr  & \Msun\ yr$^{-1}$ & 0.234$^{+0.094}_{-0.093}$ & 0.021$^{+0.012}_{-0.009}$ & 0.128$^{+0.034}_{-0.046}$ & 0.057$^{+0.034}_{-0.048}$ & 0.014$^{+0.013}_{-0.007}$ & 0.220$^{+0.093}_{-0.110}$ \\
    100-330 Myr & \Msun\ yr$^{-1}$ & 0.462$^{+0.231}_{-0.086}$ & 0.070$^{+0.015}_{-0.029}$ & 0.161$^{+0.082}_{-0.045}$ & 0.105$^{+0.035}_{-0.048}$ & 0.068$^{+0.032}_{-0.045}$ & 0.404$^{+0.164}_{-0.167}$ \\
    330 Myr-1.1 Gyr & \Msun\ yr$^{-1}$ & 0.263$^{+0.056}_{-0.085}$ & 0.065$^{+0.015}_{-0.012}$ & 0.053$^{+0.023}_{-0.027}$ & 0.044$^{+0.023}_{-0.024}$ & 0.093$^{+0.033}_{-0.024}$ & 0.255$^{+0.094}_{-0.087}$ \\
    1.1-3.6 Gyr     & \Msun\ yr$^{-1}$ & 0.239$^{+0.064}_{-0.096}$ & 0.017$^{+0.015}_{-0.010}$ & 0.066$^{+0.027}_{-0.023}$ & 0.126$^{+0.025}_{-0.029}$ & 0.030$^{+0.013}_{-0.010}$ & 0.239$^{+0.080}_{-0.072}$ \\
    3.6-11.7 Gyr     & \Msun\ yr$^{-1}$ & 0.239$^{+0.041}_{-0.034}$ & 0.100$^{+0.009}_{-0.010}$ & 0.087$^{+0.024}_{-0.023}$ & 0.023$^{+0.013}_{-0.018}$ & 0.022$^{+0.007}_{-0.005}$ & 0.232$^{+0.053}_{-0.056}$ \\
    11.7-13.7 Gyr    & \Msun\ yr$^{-1}$ & 0.222$^{+0.276}_{-0.078}$ & 0.055$^{+0.053}_{-0.043}$ & 0.261$^{+0.163}_{-0.116}$ & 0.007$^{+0.005}_{-0.009}$ & 0.032$^{+0.032}_{-0.015}$ & 0.355$^{+0.253}_{-0.183}$ \\
    \hline \hline
    \multicolumn{7}{c}{Stellar Mass Formation History}  \\
    \hline
    0-30 Myr    & log[\Msun] & 6.14$^{+0.14}_{-0.20}$ & 5.18$\pm$0.19 & 5.67$\pm$0.15 & 5.69$^{+0.13}_{-0.23}$ & 5.58$^{+0.14}_{-0.17}$ & 6.17$^{+0.14}_{-0.23}$ \\
    30-100 Myr  & log[\Msun] & 7.21$^{+0.15}_{-0.22}$ & 6.16$^{+0.20}_{-0.25}$ & 6.95$^{+0.10}_{-0.19}$ & 6.60$^{+0.26}_{-0.39}$ & 6.00$^{+0.29}_{-0.27}$ & 7.19$^{+0.24}_{-0.30}$ \\
    100-330 Myr & log[\Msun]  & 8.03$^{+0.18}_{-0.09}$ & 7.21$^{+0.09}_{-0.24}$ & 7.57$^{+0.18}_{-0.14}$ & 7.39$^{+0.16}_{-0.17}$ & 7.19$^{+0.17}_{-0.47}$ & 7.97$^{+0.16}_{-0.21}$ \\
    330 Myr-1.1 Gyr & log[\Msun] & 8.30$^{+0.08}_{-0.17}$ & 7.70$\pm$0.09 & 7.61$^{+0.16}_{-0.31}$ & 7.53$^{+0.19}_{-0.32}$ & 7.85$\pm$0.13 & 8.29$^{+0.14}_{-0.18}$ \\
    1.1-3.6 Gyr     & log[\Msun]  & 8.78$^{+0.10}_{-0.22}$ & 7.64$^{+0.28}_{-0.37}$ & 8.22$^{+0.15}_{-0.19}$ & 8.51$^{+0.09}_{-0.10}$ & 7.88$^{+0.15}_{-0.17}$ & 8.78$\pm$0.14 \\
    3.6-11.7 Gyr     & log[\Msun]  & 9.29$\pm$0.07 & 8.91$^{+0.04}_{-0.05}$ & 8.85$^{+0.10}_{-0.13}$ & 8.28$^{+0.25}_{-0.36}$ & 8.26$^{+0.09}_{-0.16}$ & 9.07$^{+0.09}_{-0.10}$ \\
    11.7-13.7 Gyr    & log[\Msun]  & 8.66$_{-0.35}^{+0.19}$ & 8.05$^{+0.29}_{-0.68}$ & 8.73$^{+0.21}_{-0.26}$ & 7.17$_{-0.35}^{+0.61}$ & 7.82$^{+0.30}_{-0.26}$ & 8.86$^{+0.25}_{-0.30}$
    \enddata
\end{deluxetable*}

\vspace*{-\baselineskip}\vspace*{-\baselineskip}

\begin{figure}
    \centering
     \includegraphics[width=1\linewidth]{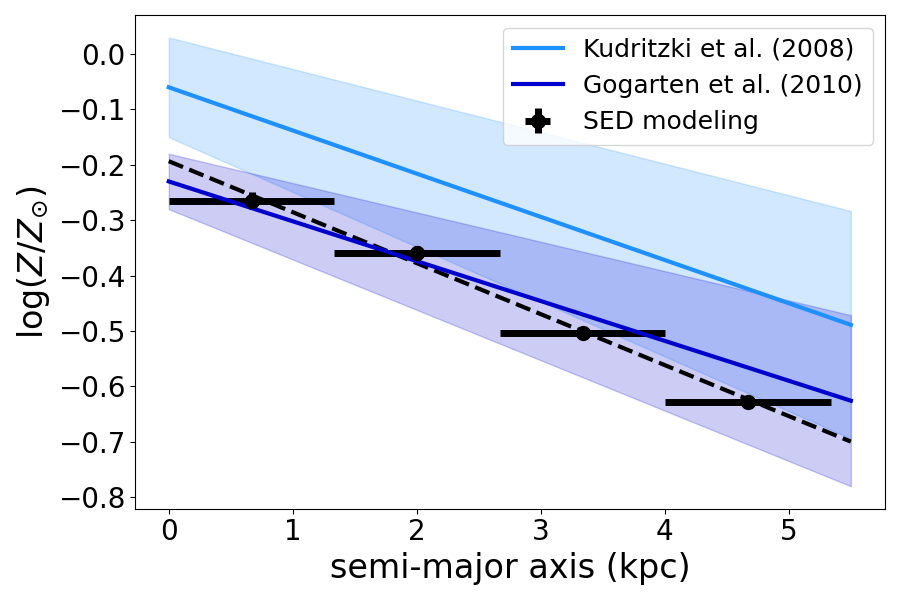} \\
     \includegraphics[width=1\linewidth]{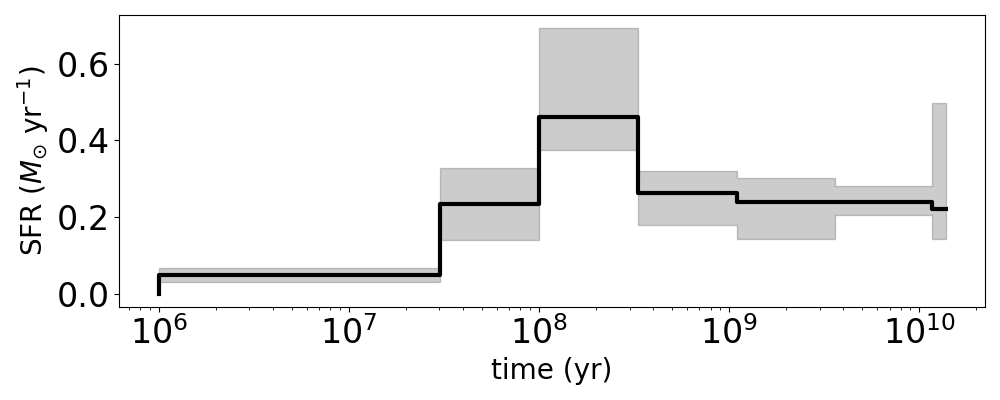} \\
    \caption{{\it Top}: The metallicity gradient of NGC~300. The black points show the results of our SED modeling with \prospector, and the dashed black line shows the best-fit metallicity gradient. The metallicity gradients taken from the literature (and their corresponding uncertainties) are shown in light blue \citep{Kudritzki+08} and dark blue \citep{Gogarten+10}. {\it Bottom}: The SFH of NGC~300 (black; gray shaded region indicates uncertainties). }
    \label{fig:metallicity_gradient}
\end{figure}

The metallicity gradient recovered from the \prospector\ is shown in the top panel of Figure~\ref{fig:metallicity_gradient} compared to the gradients of \citet[][light blue]{Kudritzki+08} and \citet[][dark blue]{Gogarten+10}. Our SED modeling of NGC~300 yields a metallicity gradient of

\begin{equation}
    \text{log}\left(Z/Z_{\odot}\right) = \left(-0.193\pm0.016 \right) - \left(0.092\pm0.005 \right)r,
\end{equation}

\noindent where $r$ is the galactocentric radius in kpc. This gradient shows good agreement with those taken from the literature and predicts a central metallicity value that is intermediate between the \citet{Gogarten+10} and \citet{Kudritzki+08} values. Our utilization of \prospector\ is similar to CMD fitting in that we rely on similar theoretical stellar models to construct galaxy SEDs for comparison to flux densities, rather than directly fitting for abundances from spectra. We note that the radial gradient obtained by \citet{Gogarten+10} was derived using a narrow strip of three \hst\ fields that spanned the disk of NGC~300 (which also prohibited measurement of integrated galaxy mass and SFR), while our SED modeling uses observations extending over the full annular rings shown in Figure~\ref{fig:RGB}. The metallicity gradient of NGC~300 from \citet{Kudritzki+08} was derived from a sample of 24 supergiants (with spectral types ranging from B8 to A4), and \citet{Bresolin+09} found that these direct stellar metallicity measurements were in excellent agreement with those derived from a sample of 28 H~II regions for which auroral lines were detected. The stellar and auroral nebular metallicities did not, however, agree with abundances derived from strong-line methods: the metallicity gradient based on strong oxygen lines in \citet{Zaritsky+94} and \citet{Deharveng+88} $\sim$50\% steeper than that found in \citet{Kudritzki+08} and \citet{Bresolin+09}. The metallicities of H~II regions in two other well-studied spiral galaxies – the Milky Way and M33 – show significant discrepancies depending on the lines and methods used, and these differences cannot be explained by measurement errors. Using mid- and far-IR lines largely resolves the discrepancies in the Milky Way \citep{MartinHernandez+02,Rudolph+06}, bringing the nebular metallicity gradient into agreement with that of direct stellar measurements, but these lines are rarely used to derive metallicities or abundance gradients in external spiral galaxies. Additional discussion of metallicity and its role in HMXB evolution is discussed in Section~\ref{sec:discussion}.

\begin{figure*}
    \centering
    \includegraphics[width=1\linewidth]{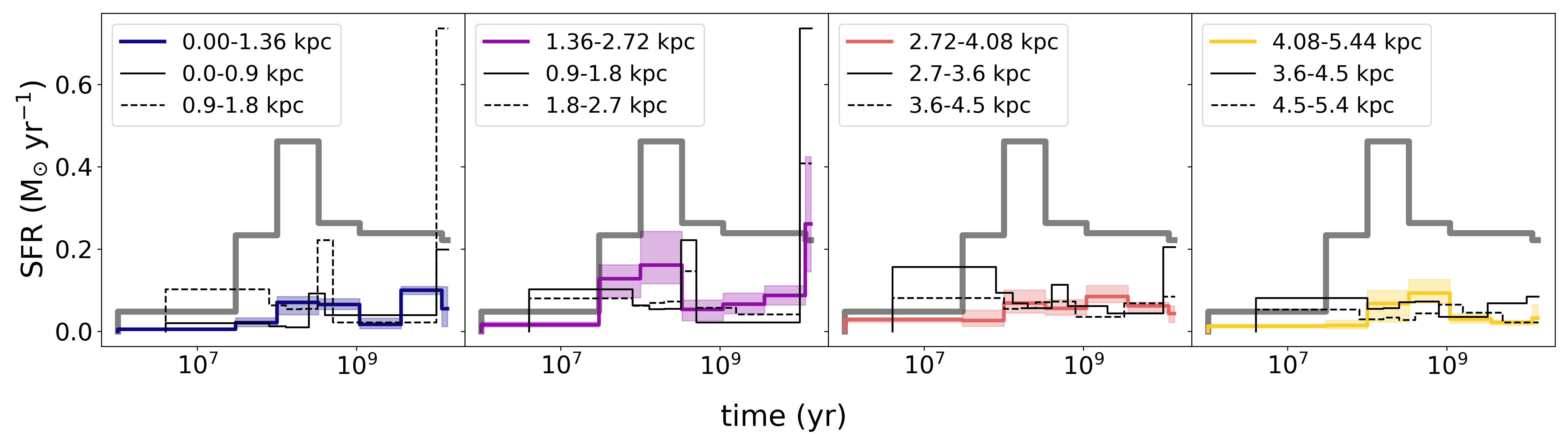}
    \caption{Radially-resolved SFHs derived from \prospector\ modeling (color-coded as in Figure~\ref{fig:SEDs}) with increasing galactocentric radius (from left to right) compared to the SFHs derived by \citet[][solid and dashed black lines]{Gogarten+10} at similar galactocentric distances. The SFHs from \citet{Gogarten+10} carry a typical uncertainty of $\sim$10-20\% and were computed over a much smaller surface area of NGC~300 than the \prospector-derived SFHs. The galaxy-integrated SFH from Figure~\ref{fig:metallicity_gradient} is shown by the thick gray line.}
    \label{fig:SFH_comparison}
\end{figure*}

The overall SFH for the entire galaxy (Figure~\ref{fig:metallicity_gradient}, bottom panel) is in broad agreement with previous studies, showing an enhancement in the SFR at early times \citep[roughly 90\% the stars ever formed in NGC~300 were formed more than a few Gyr ago, as found by][]{Gogarten+10}. The average SFR on timescales relevant for HMXBs ($<$100 Myr) agrees with estimates derived by other SFR indicators, our SED modeling suggests that most of the star formation activity occurred $\sim$30-100 Myr ago, with the most recent SFR ($<$30 Myr) occurring at only $\sim$20\% the rate found for the older time bin. \citet{Williams+13} found that most HMXBs in NGC~300 are $\sim$40-60 Myr old through CMD analysis of the resolved stellar populations in the vicinity of HMXB candidates, with NGC~300 ULX-1 being the only notable exception of a younger HMXB \citep[the massive HMXB NGC~300 X-1 is likely $<$30 Myr as well, given the large BH mass and evolved state of the Wolf-Rayet companion;][]{Binder+20,Binder+21}. We next compared the SFHs derived by \prospector\ to the radially-resolved SFHs of \citet{Gogarten+10}, which assumed a \citet{Salpeter55} IMF during CMD fitting but report results scaled to a \citet{Kroupa01} IMF. Although the outermost radii ($\sim$5.4 kpc) are similar in both studies, we use four bins each with a width of 1.36 kpc while \citet{Gogarten+10} used six bins of width 0.9 kpc, and the time bins used for in constructing the SFHs differs. Furthermore, the \citet{Gogarten+10} analysis was limited to the small area covered by the \hst\ fields, whereas our SED modeling is performed for annular rings across the entire disk. This introduces some additional uncertainty in the comparison of the two radially-resolved SFHs. The SFHs show broad agreement (uncertainties in the \citet{Gogarten+10} SFRs are typically on the order of $\sim$20\%; Figure~\ref{fig:SFH_comparison}), with the largest discrepancies appearing in the oldest time bin at $r < 2R_d$, where \citet{Gogarten+10} predict a higher SFR at very early times compared to our \prospector\ results, particularly at $r\lesssim$4 kpc. These differences will not affect our interpretation of the HMXB population properties, which depend only on the most recent SFRs. 

Different SFR indicators are sensitive to different population ages, and \citet{Kouroumpatzakis+20} demonstrated that the relationship between \Lx\ and SFR of HMXB populations depends on choice of SFR indicator. As shown in Table~\ref{tab:params_lit}, different SFR indicators can yield up to a factor of $\sim$5 difference in inferred SFR in NGC~300 on a galaxy-integrated scale. Metallicity, too, can be measured through a variety of means; nearly all studies of the \Lx-SFR-$Z$ relationship assume the the overall metallicity of a galaxy is well traced by the nebular oxygen abundance, 12 + log(O/H). While emission lines from H~II region spectra provide the bulk of direct present-day abundance measurements, many studies utilize metallicities derived from strong-line indices such as [O III] $\lambda$5007. Auroral emission lines that directly probe both the composition and the electron temperatures of H~II regions (e.g., [O~III] $\lambda$4363) become fainter with both increasing distance and metallicity (as the electron temperature decreases due to cooling via metal lines; \citep{Berg+15,Croxall+15,Croxall+16}, eventually becoming to faint to detect. Strong-line metallicity indices therefore depend either on empirical calibrations tied to more local auroral line measurements \citep{Pettini+04,Pilyugin+05,Liang+07,Florido+22} or theoretical calibrations derived from grids of photoionization models \citep{McGaugh91,Kewley+02,Tremonti+04} These different calibrations provide large differences ($\sim$0.7 dex) in the inferred metallicities \citep{Kewley+08}, and may be significantly contributing to the observed scatter in the \Lx-SFR-$Z$ relationships.

\section{X-ray Observations from Chandra} \label{sec:data_xray}
We retrieved four \chandra\ ACIS-I observations (ObsIDs 12238, 16028, 16029, and 22375) from the \chandra\ archive\footnote{\url{https://cda.harvard.edu/chaser/}}. The observations ranged in exposure duration from 47.4 ks (ObsID 22375) to 64.2 ks (ObsID 16028). We reprocessed all \texttt{evt1} data using the CIAO v4.15 \citep{Fruscione+06} task \texttt{chandra\_repro} and standard reduction procedures. We ran the point source detection algorithm \texttt{wavdetect} on each individual exposure to generate a preliminary list of X-ray source positions and error ellipses. The major and minor axes of these error ellipses were increased by a factor of five, and all X-ray sources were masked so that a background light curve could be extracted for each observation. We inspected these background light curves but found no evidence for strong, prolonged background flaring events in any of the four observations. We created good time intervals (GTIs) using \texttt{lc\_clean} and filtered all event data on these GTIs. Finally, we restricted the energy range of our resulting ``clean'' \texttt{evt2} files to 0.5-7 keV. 

We updated the absolute astrometry of all four images by using the tasks \texttt{wcs\_match} and \texttt{wcs\_match} to align each \chandra\ observation to the longest observation (ObsID 16028). Images were finally reprojected to a common tangent point and combined using the CIAO task \texttt{reproject\_obs} and \texttt{flux\_obs}. The total usable \chandra\ exposure time was 235.9 ks. We next re-ran \texttt{wavdetect} on the merged observation, using smoothing scales of ``1 2 4 8'' and a significance threshold of $10^{-6}$. A total of 154 X-ray sources were detected in the merged \chandra\ image. An RGB-rendered image of NGC~300 with the locations of all \chandra\ X-ray sources is shown in Figure~\ref{fig:RGB}. A summary of the properties of all the detected X-ray sources is provided in Table~\ref{tab:X-ray_srclist} and provided as a machine-readable table from the journal. 

\begin{deluxetable*}{ccccccccccccccccc}
\tabletypesize{\scriptsize}
\setlength{\tabcolsep}{1pt}
    \caption{Properties of All 154 X-ray Sources Detected in Merged \chandra\ Image\tablenotemark{a}}\label{tab:X-ray_srclist}
    \tablehead{
     \colhead{RA} & \colhead{Dec.} & \colhead{Galactocentric}  & \colhead{Galactocentric} & \colhead{Source}  & \colhead{Net Counts} & \multicolumn{2}{c}{12238} && \multicolumn{2}{c}{16028} && \multicolumn{2}{c}{16029} && \multicolumn{2}{c}{22375}  \\ \cline{1-2} \cline{7-8} \cline{10-11} \cline{13-14} \cline{16-17}
     \multicolumn{2}{c}{(J2000)}        & Radius (kpc) & Radius/$R_d$ & Signif. ($\sigma$) & (0.5-7 keV) & HR1 & HR2 && HR1 & HR2 && HR1 & HR2 && HR1 & HR2 \\
     }
    \startdata
13.6980206 & -37.8927248 & 10.10 & 7.43 & 20.5 & 549$\pm$35 & \nodata & \nodata && \nodata & \nodata && \nodata & \nodata && 0.42$^{+0.05}_{-0.05}$ & -0.20$^{+0.06}_{-0.06}$ \\
13.7722944 & -37.8583461 & 8.07 & 5.93 & 4.2 & 35$\pm$9 & \nodata & \nodata && \nodata & \nodata && 0.75$^{+0.12}_{-0.12}$ & -0.60$^{+0.21}_{-0.22}$ && \nodata & \nodata \\
13.6859116 & -37.8505873 & 8.22 & 6.04 & 13.9 & 179$\pm$18 & \nodata & \nodata && 0.47$^{+0.10}_{-0.10}$ & -0.33$^{+0.14}_{-0.15}$ && 0.44$^{+0.08}_{-0.08}$ & -0.33$^{+0.10}_{-0.11}$ && \nodata & \nodata \\
13.6999274 & -37.8355104 & 7.37 & 5.42 & 11.3 & 106$\pm$13 & \nodata & \nodata && 0.43$^{+0.15}_{-0.14}$ & -0.31$^{+0.19}_{-0.22}$ && 0.40$^{+0.10}_{-0.10}$ & -0.15$^{+0.13}_{-0.15}$ && \nodata & \nodata \\
13.7920276 & -37.8333377 & 6.91 & 5.08 & 7.2 & 68$\pm$12 & \nodata & \nodata && \nodata & \nodata && 0.76$^{+0.15}_{-0.19}$ & -0.78$^{+0.23}_{-0.14}$ && 0.28$^{+0.21}_{-0.21}$ & -0.24$^{+0.27}_{-0.30}$ \\
13.8005427 & -37.8284841 & 6.71 & 4.93 & 16.1 & 177$\pm$17 & \nodata & \nodata && \nodata & \nodata && 0.53$^{+0.09}_{-0.09}$ & -0.30$^{+0.13}_{-0.14}$ && 0.47$^{+0.08}_{-0.08}$ & -0.08$^{+0.13}_{-0.14}$ \\
13.7174112 & -37.8202939 & 6.50 & 4.78 & 5.6 & 49$\pm$10 & \nodata & \nodata && 0.41$^{+0.22}_{-0.21}$ & -0.18$^{+0.30}_{-0.36}$ && 0.56$^{+0.22}_{-0.22}$ & -0.44$^{+0.35}_{-0.32}$ && 0.22$^{+0.49}_{-0.48}$ & -0.08$^{+0.65}_{-0.63}$ \\
13.6970791 & -37.8204055 & 6.68 & 4.91 & 4.4 & 35$\pm$9 & \nodata & \nodata && 0.45$^{+0.19}_{-0.16}$ & -0.01$^{+0.31}_{-0.37}$ && 0.70$^{+0.17}_{-0.18}$ & -0.58$^{+0.29}_{-0.26}$ && \nodata & \nodata \\
13.7205749 & -37.8120096 & 6.09 & 4.48 & 4.6 & 53$\pm$13 & \nodata & \nodata && 0.52$^{+0.20}_{-0.19}$ & -0.53$^{+0.26}_{-0.23}$ && 0.44$^{+0.26}_{-0.28}$ & -0.29$^{+0.40}_{-0.40}$ && -0.01$^{+0.53}_{-0.54}$ & -0.34$^{+0.34}_{-0.31}$ \\
13.7953472 & -37.8095754 & 5.85 & 4.30 & 34.4 & 416$\pm$23 & \nodata & \nodata && 0.42$^{+0.11}_{-0.11}$ & -0.53$^{+0.13}_{-0.13}$ && 0.36$^{+0.06}_{-0.07}$ & -0.36$^{+0.07}_{-0.07}$ && 0.51$^{+0.06}_{-0.06}$ & -0.33$^{+0.08}_{-0.08}$ 
    \enddata
    \tablenotetext{a}{Only the first ten entries are shown. A full machine-readable table is available from the journal.}
\end{deluxetable*}

\begin{figure}
    \centering
    \includegraphics[width=1\linewidth]{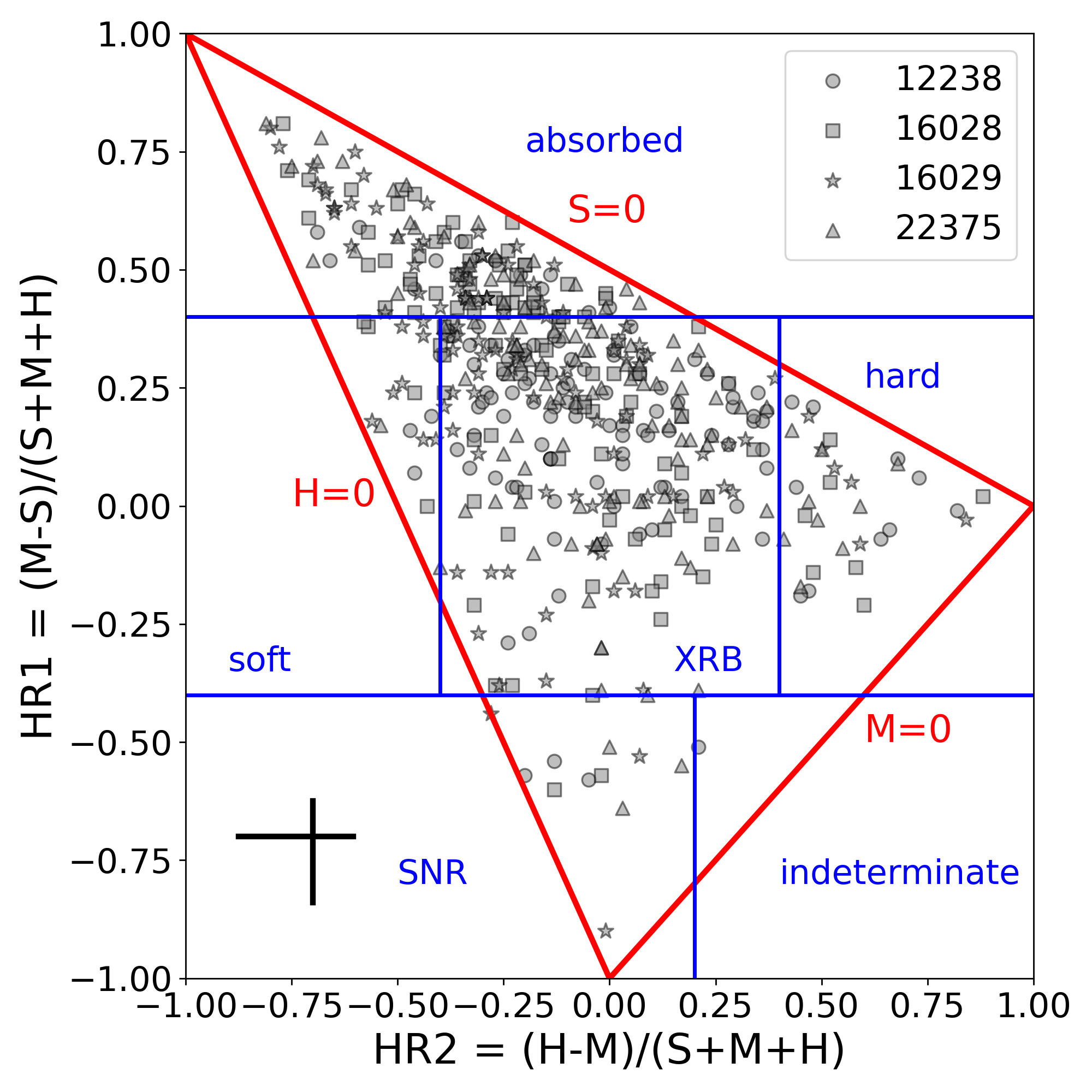}
    \caption{Hardness ratio diagram for all 154 sources detected in the merged \chandra\ image of NGC~300, with the six hardness ratio categories defined in \citet{Binder+12} shown in blue. Red lines indicate the physically meaningful region of the diagram. The typical uncertainty on the hardness ratios is shown by the black cross in the lower left corner. }
    \label{fig:hardness_ratios}
\end{figure}

We additionally calculate hardness ratios for each X-ray source using the Bayesian estimation of hardness ratios (BEHR) code \citep{Park+06} following the same methodology as in \citet{Binder+12}. We define two hardness ratios,

\begin{equation}
    \text{HR1} = \frac{M-S}{H+M+S},
\end{equation}

\noindent and

\begin{equation}
    \text{HR2} = \frac{H-M}{H+M+S},
\end{equation}

\noindent where $S$ is the ``soft'' band (0.35-1.1 keV), $M$ is the ``medium'' band (1.1-2.6 keV), and $H$ is the ``hard'' band (2.6-8.0 keV). We use BEHR to generate 50,000 samples of the source flux probability distributions (in all four observations) for each of the three bands, which we use to compute 50,000 values of HR1 and HR2 for each source. In Table~\ref{tab:X-ray_srclist} we report the median value for each hardness ratio, and the uncertainties are defined using the 16th and 84th percentile values. We use the hardness ratio categories defined in \citet{Binder+12} to roughly classify sources as likely XRBs, SNRs, highly absorbed sources (``ABS''), indeterminate soft or hard sources (``SOFT'' and ``HARD,'' respectively), and sources with indeterminate spectral types (``INDET''). The hardness ratio diagram in Figure~\ref{fig:hardness_ratios} shows the hardness ratios for each source and the six categories defined in \citet{Binder+12}.

\subsection{Sensitivity Curves and Limiting Fluxes}\label{sec:sensitivity}
In addition to generating a merged observation image, we also created background images using the CIAO \texttt{blank\_sky} task. ``Blank sky'' images were generated to estimate the source-free background across the detector for each observation and merged in the same manner as described above for the actual observation images. These blank sky images can be used to derive sensitivity curves, which provide an estimate of the probability that an X-ray source with a given flux will be detected at a given location on the detector. These are essential components for accurately modeling XLFs and for determining the flux levels below which a source catalog is highly incomplete.

We follow the methodology of \citet{Georgakakis+08} to construct a sensitivity curve for the merged \chandra\ image of NGC~300. We present a summary of the sensitivity curve construction procedure here; the reader is referred to \citet{Georgakakis+08} for additional details and an example application to AGN observations. The merged blank sky image provides us with an estimate of the mean expected source-free background counts $B$ across the entire image. The cumulative probability that the observed number of counts detected in a source (which is a sum of the counts intrinsic to the source and the background) will exceed some limit $L$ is given by

\begin{equation}\label{eq:sensitivity_map_eq}
    P_B(\geq L) = \frac{1}{\Gamma(L)} \int_0^B e^{-t} t^{L-1} dt
\end{equation}

\noindent assuming $B$ and $L$ are integers. If one adopts a value for the threshold probability $P_B$, Equation~\ref{eq:sensitivity_map_eq} can be inverted to find the detection limit $L$ for a given background. We adopt a threshold probability of $P_B=4\times10^{-6}$ \citep[the values assumed by, e.g., ][which corresponds to approximately 0.5 false sources per \chandra\ image]{Georgakakis+08,Laird+09} and perform this calculation across the entire merged image area to construct a map of $L$ values as a function of ($x,y$), given the background counts $B$ predicted by the blank sky map. We can convert $L$ to an energy flux by assuming a power law spectral model (with photon index $\Gamma=1.7$ and subject to the Galactic absorbing column) and using the total exposure time at each ($x,y$) position across the image. From the resulting 2D fluxed image, we measure the total area over which a source with a given flux could be detected. 

We perform this calculation for the whole galaxy out to a maximum distance of $4R_d$ (we note that beyond this radius the X-ray point source population is likely to be both highly incomplete due to the lower sensitivity at large off-axis angle and dominated by background AGN, rather than sources intrinsic to NGC~300). The 90\% limiting 0.5-7 keV flux is 3.38$\times10^{-15}$ \flux\ \citep[corresponding to a luminosity of 1.6$\times10^{36}$ \lum\ at the distance of NGC~300, in good agreement with][who used a subset of the observations analyzed in this work but a somewhat less restrictive threshold for detection]{Binder+12,Binder+17}. We hereafter refer to the 90\% limiting luminosity as $L_{\rm X,90}$ and the 90\% limiting flux as $S_{\rm X,90}$.

For the remainder of our analysis, we only keep X-ray sources that meet all of the following three criteria: (1) detected at $r\leq 4R_d$, (2) detected at $\geq3\sigma$ significance by \texttt{wavdetect}, and (3) with fluxes $\geq S_{\rm X,90}$. Within $4R_d$, there is one supernova remnant \citep{Gross+19} and one foreground star \citep{Binder+12} that we remove from the X-ray source list. A total of 70 X-ray sources meet these criteria, with 6 located in the inner region, 16 in the inner middle bin, 22 in the outer middle bin, and 26 in the outer bin. These sources are indicated in yellow on Figure~\ref{fig:RGB}.

\subsection{The X-ray Binary Population}\label{sec:XRBs}
To estimate the number and total \Lx\ produced by XRBs in NGC~300, we first used the AGN log$N$-log$S$ distribution from the \chandra\ Multiwavelength Project \citep[ChaMP,][]{Kim+04} to estimate the AGN contribution in each annular region. We estimate 43$\pm$2 AGN above $S_{\rm X,90}$ are found within the full survey area ($\leq4R_d$), and $\sim$7 of these are AGN with fluxes above $\sim2.1\times10^{-14}$ \flux\ (which corresponds to a luminosity of $10^{37}$ \lum\ at the distance of NGC~300). The remaining $\sim$27 X-ray sources above $S_{\rm X,90}$ ($\sim$6 above $10^{37}$ \lum) are therefore likely intrinsic to NGC~300. 

When we use the area spanned by each elliptical annulus, the predicted number of AGN above $S_{\rm X,90}$ in each subgalactic radial bin (from the innermost region outward) is approximately 3, 8, 14, and 18. Given their far distance, background AGN are more likely to be highly absorbed than XRBs and other X-ray sources intrinsic to NGC~300; we therefore examine the number of sources with hardness ratios placing them in the ``ABS'' category of Figure~\ref{fig:hardness_ratios} in at least one observation. We find 3, 13, 16, 20 sources show evidence of high absorption in each of the four radial bins. This is only slightly higher than the predicted number of background AGN, and consistent with expectations given that some XRBs can experience significant intrinsic absorption. We therefore expect approximately 3, 8, 8, and 8 X-ray sources above $S_{X,90}$ in the inner, inner middle, outer middle, and outer bin, respectively, are XRBs in NGC~300. We additionally measure the total flux we expect to originate from AGN and subtract this value from the flux of the total X-ray source population; the remainder is the approximate flux of XRBs in NGC~300. We use the reported uncertainties on the ChaMP log$N$-log$S$ distribution to assess the uncertainty in the number of AGN predicted in our observations; the uncertainty is $\sim$2-3 AGN for the whole galaxy. These calculations are summarized in Table~\ref{tab:Xray_sources}.

Given the relatively low number of likely XRBs observed in each radial bin, we do not attempt to directly fit the XRB X-ray luminosity functions. Instead, we predict the expected number of HMXBs and LMXBs and their total \Lx\ using the (metallicity-independent) HMXB and LMXB scaling relations of \citet{Lehmer+20,Lehmer+21}: 

\begin{equation}
  \label{eq:scaling_relations}
  \begin{aligned}
    \text{log}\left(\frac{L_{\rm X}^{\rm HMXB}}{\text{SFR}} \right) &= 39.71^{+0.14}_{-0.09} \left[\frac{\text{erg s}^{-1}}{M_{\odot} \text{ yr}^{-1}} \right] \\     
    \text{log}\left(\frac{L_{\rm X}^{\rm LMXB}}{M_{\star}} \right) &= 29.25^{+0.07}_{-0.06} \left[\frac{\text{erg s}^{-1}}{M_{\odot}} \right]
  \end{aligned}
\end{equation}

\noindent These scaling relations were derived from a sample of 24 elliptical galaxies \citep{Lehmer+20} and 33 star-forming (average SFR$\sim$1.4 \Msun), HMXB-dominated galaxies \citep{Lehmer+21}. The XLFs of some of these galaxies extend below 10$^{37}$ \lum\ \citep[4/22 elliptical galaxies had 90\% limiting luminosities below $10^{37}$ \lum\ and 9/33 star-forming galaxies had 50\% limiting luminosities comparable to $L_{\rm X,90}$ in this work;][]{Lehmer+20,Lehmer+21}, however the majority of the X-ray sources used to derive these scaling relations were significantly brighter than most of the X-ray sources coincident with the disk of NGC~300. Given the relatively small number of galaxies with X-ray source populations imaged to faint fluxes, the applicability of these scaling relations to faint XRB populations may be limited (see Section~\ref{sec:discussion} for further discussion).

Since the SFRs and stellar masses considered here are low and XRB formation is a stochastic process, we use the following approach to estimate the XRB population properties expected for NGC~300. First, we randomly draw values of the SFR and \Mstar\ from within the uncertainties derived from our SED modeling for each subgalactic region in NGC~300 assuming a normal distribution with the mean set equal to the best-fit value of each parameter. Since our best-fit values of the SFR and \Mstar\ have asymmetric errors, we set the standard deviation of this normal distribution to the larger uncertainty. We then randomly select a value from within the uncertainties of the above scaling relations and compute the expected \Lx\ for each population. To convert the predicted total \Lx\ into a number of XRBs, we assume that both HMXBs and LMXBs follow the broken power law models described in \citet{Lehmer+19,Lehmer+20}. 

To randomly generate a sample of synthetic XRBs that follow a broken power law distribution, we follow the approach outlined in \citet[][Appendix D]{Clauset+09}. We first generate a random \Lx\ value for an XRB according to $L_{\rm X,r}=L_{\rm X, 90}(1-r)^{-1/(\gamma_1-1)}$, where $\gamma_1$ is the faint-end power law slope and $r$ is a number that is randomly drawn from a uniform distribution in the range $0\leq r < 1$. If $L_{\rm X,r}$ is brighter than the 10$^{39}$ \lum\ cutoff luminosity assumed in \citet{Lehmer+19,Lehmer+20} it is rejected. If $L_{\rm X,r}$ is within the observed range of \Lx\ values as the true sample, we draw a second uniform random variate $r_2$ (also in the range $0\leq r_2<1$) and accept $L_{\rm X,r}$ if $(L_{\rm X,r}/L_{\rm X,90})^{-\gamma_1} \geq r_2$. If $L_{\rm X,r}$ is rejected, we use a similar functional form as above to generate a new value of $L_{\rm X,r}$ that follows the bright end of the XLF: $L_{\rm X,r}=L_{\rm X, break}(1-r)^{-1/(\gamma_2-1)}$, where $\gamma_2$ is the slope of the bright end of the XLF and $L_{\rm X, break}$ is the break luminosity \citep[$3\times10^{37}$ \lum\ for LMXBs and $10^{38}$ \lum\ for HMXBs;][]{Lehmer+19,Lehmer+20}. Synthetic XRBs are generated one at a time and their total luminosity is tracked until the value predicted using the scaling relations and relevant SFR and \Mstar\ values is obtained. This process is repeated 10\,000 times to build up a distribution of synthetic HMXB and LMXB populations that satisfy the scaling relations in Equation~\ref{eq:scaling_relations} for the subgalactic regions of NGC~300. We used the median and 90\% confidence interval to define the ``predicted'' number of XRBs and their collective \Lx\ and report these values in Table~\ref{tab:Xray_sources}. For clarity, we use use $N$ and \Lx\ to refer to numbers of X-ray sources and their luminosity measured or inferred from the data, respectively, and $\mathcal{N}$ and $\mathcal{L}_X$ are used to indicate numbers and luminosities, respectively, that are predicted purely from the XRB scaling relations as described above.

\begin{deluxetable*}{ccccccc}
    \caption{The Predicted and Inferred X-ray Source Population in NGC~300 }\label{tab:Xray_sources}
    \tablehead{
    & Galaxy Inte-  & \multicolumn{4}{c}{Radial Bin} & Sum of All \\ \cline{3-6}
     & \colhead{grated ($r\leq 4R_d$)} & \colhead{$r\leq R_d$} & \colhead{$R_d < r \leq 2R_d$}  & \colhead{$2R_d < r \leq 3R_d$}  & \colhead{$3R_d < r \leq 4R_d$}  & Radial Bins
     }
    \colnumbers
    \startdata
    \multicolumn{7}{c}{Observed X-ray Point Sources} \\
    \hline
    $N_{\rm X}$   & 70 & 6 & 16 & 22 & 26 & \nodata \\
    $N_{\rm AGN}$\tablenotemark{a} ($>S_{\rm X,90}$) & 42.8$\pm$2.1 & 2.8$\pm$0.1 & 8.0$\pm$0.4 & 14.1$\pm$0.8 & 17.9$\pm$0.9 & \nodata \\
    $N_{\rm AGN}^{\rm HR}$\tablenotemark{b} & 52 & 3 & 13 & 16 & 20 & \nodata \\
    $N_{\rm XRB}^{\rm HR}$\tablenotemark{b} & 18 & 3 & 3 & 6 & 6 & \nodata \\
    $N_{\rm int}$ ($>L_{\rm X,90}$) & 27.2$\pm$2.1 & 3.2$\pm$0.1 & 8.0$\pm$0.4 & 7.9$\pm$0.8 & 8.1$\pm$0.9 & \nodata \\
    $N_{\rm int}$ ($>10^{37}$ \lum) & 6.4$\pm$0.4 & 0.6$\pm$0.1 & 2.8$\pm$0.1 & 3.0$\pm$0.1 & $<$0.2 & \nodata \\
    $L_{\rm X, int}$\tablenotemark{d} (10$^{38}$ \lum)  & 4.9$\pm$0.4 & 0.16$\pm$0.03 & 3.9$\pm$0.3 & 0.4$\pm$0.1 & 0.4$\pm$0.2 & \nodata  \\
    \hline
    \multicolumn{7}{c}{Predicted X-ray Point Sources\tablenotemark{c}} \\
    \hline
    $\mathcal{N}_{\rm HMXB}$ ($>L_{\rm X,90}$) & 4$^{+2}_{-3}$ & 1$\pm$1 & 3$^{+2}_{-1}$ & 2$^{+2}_{-1}$ & 1$\pm$1 & 7$^{+6}_{-4}$ \\
    $\mathcal{N}_{\rm HMXB}$ ($>10^{37}$ \lum) & 4$^{+2}_{-1}$ & 1$\pm$1 & 2$^{+2}_{-1}$ & 2$\pm$1 & 1$\pm$1 & 6$^{+5}_{-4}$ \\
    $\mathcal{N}_{\rm LMXB}$ ($>L_{\rm X,90}$) & 3$\pm$2 & 1$^{+2}_{-1}$ & 2$\pm$1 & 2$^{+2}_{-1}$ & 1$\pm$1 & 6$^{+6}_{-4}$ \\
    $\mathcal{N}_{\rm LMXB}$ ($>10^{37}$ \lum) & 2$^{+2}_{-1}$ & 1$\pm$1 & 1$^{+2}_{-1}$ & 1$\pm$1 & 1$\pm$1 & 4$^{+5}_{-4}$ \\
    $\mathcal{N}_{\rm HMXB+LMXB}$ ($>L_{\rm X,90}$) & 7$^{+4}_{-5}$ & 2$^{+3}_{-2}$ & 5$^{+3}_{-2}$ & 4$^{+4}_{-2}$ & 2$\pm$2 & 13$^{+12}_{-8}$ \\
    $\mathcal{L}_X$(HMXB) ($10^{38}$ \lum)   & 7.6$^{+2.6}_{-2.2}$ & 0.6$\pm$0.3 & 4.3$^{+1.3}_{-1.1}$ & 2.6$^{+1.2}_{-1.1}$ & 0.6$^{+0.3}_{-0.2}$ & 8.1$^{+3.1}_{-2.7}$ \\
    $\mathcal{L}_X$(LMXB) ($10^{38}$ \lum)   & 4.5$^{+0.5}_{-0.4}$ & 1.0$\pm$0.2 & 1.6$\pm$0.2 & 0.7$\pm$0.1 & 0.5$\pm$0.1 & 3.8$\pm$0.6 \\
    $\mathcal{L}_X^{\rm HMXB+LMXB}$ ($10^{38}$ \lum) & 12.1$^{+3.1}_{-2.6}$ & 1.6$\pm$0.5 & 5.9$^{+1.5}_{-1.3}$ & 3.3$^{+1.3}_{-1.2}$ & 1.1$^{+0.4}_{-0.3}$ & 11.9$^{+3.7}_{-3.3}$
    \enddata
    \tablerefs{$^a$Inferred from the AGN log$N$-log$S$ distribution \citep{Kim+04}. $^b$Number of AGN and XRB candidates inferred from hardness ratio analysis. $^c$Values are not corrected for the true \Lx\ of NGC~300 ULX-1 (see Section~\ref{sec:obs_vs_pred} for details); when corrected for NGC~300 ULX-1, the galaxy-integrated $L_{\rm X, int}$ is 4.43$\times10^{39}$ \lum\ and the corresponding radial bin ($R_d<r \leq 2R_d$) $L_{\rm X, int}$ becomes 4.36$\times10^{39}$ \lum. \\
    Subgalactic regions are modeled independently; the sums in columns (3)-(6) agree (within the uncertainties) with the spatially-integrated results in column (2).}
\end{deluxetable*}

\vspace{-0.5cm}
\subsection{Observed vs. Predicted X-ray Source Populations}\label{sec:obs_vs_pred}
\begin{figure}
    \centering
    \includegraphics[width=1\linewidth]{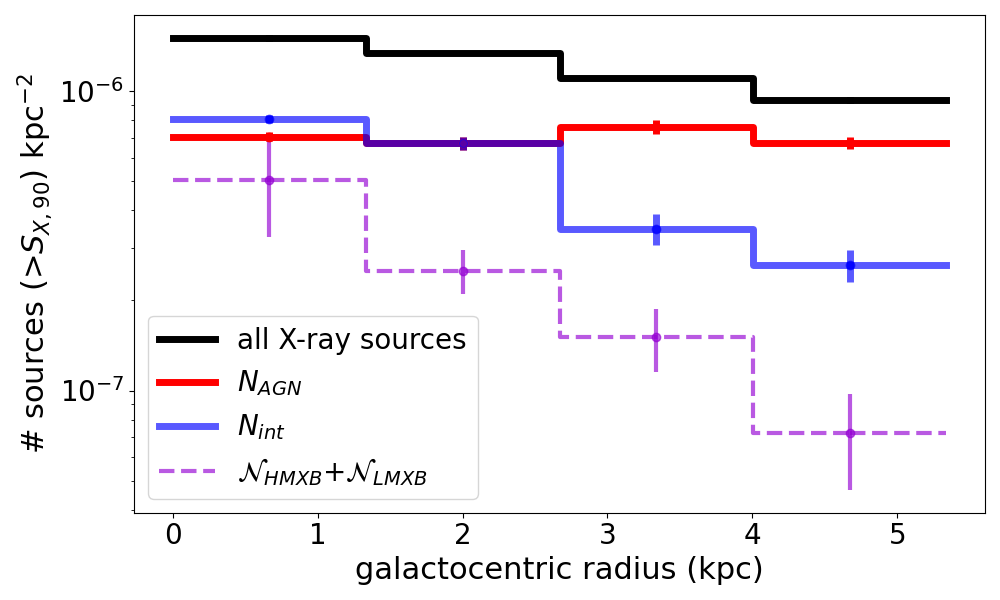}
    \caption{Radial distribution of all X-ray sources (black), the AGN contribution inferred from the ChaMP log$N$-log$S$ distribution (red), the observed X-ray source population expected to be intrinsic to NGC~300 (blue), and the combined HMXB and LMXB population expected from the XRB scaling relations (dashed purple). }
    \label{fig:radial_distr}
\end{figure}

We observe more X-ray sources above $S_{\rm X,90}$ than predicted for the field of NGC~300 based on the ChaMP AGN log$N$-log$S$ distribution and the scaling relations of \citet{Lehmer+20,Lehmer+21}. Figure~\ref{fig:radial_distr} shows the observed radial source distribution (black), the AGN contribution inferred from the ChaMP log$N$-log$S$ distribution (red), the observed X-ray source population expected to be intrinsic to NGC~300 (blue), and the combined HMXB and LMXB population expected from the XRB scaling relations (dashed purple) above $S_{\rm X,90}$. Only in the central region ($r\leq R_d$) of NGC~300 is there near-consistency between observations and predictions (owing to the large uncertainties in the expected number of XRBs), while the observed number of X-ray sources in the outer three regions exceeds predictions by $\ge 3\sigma$.

\begin{figure}
    \centering
    \includegraphics[width=1\linewidth,trim={0 2.6cm 0 0},clip]{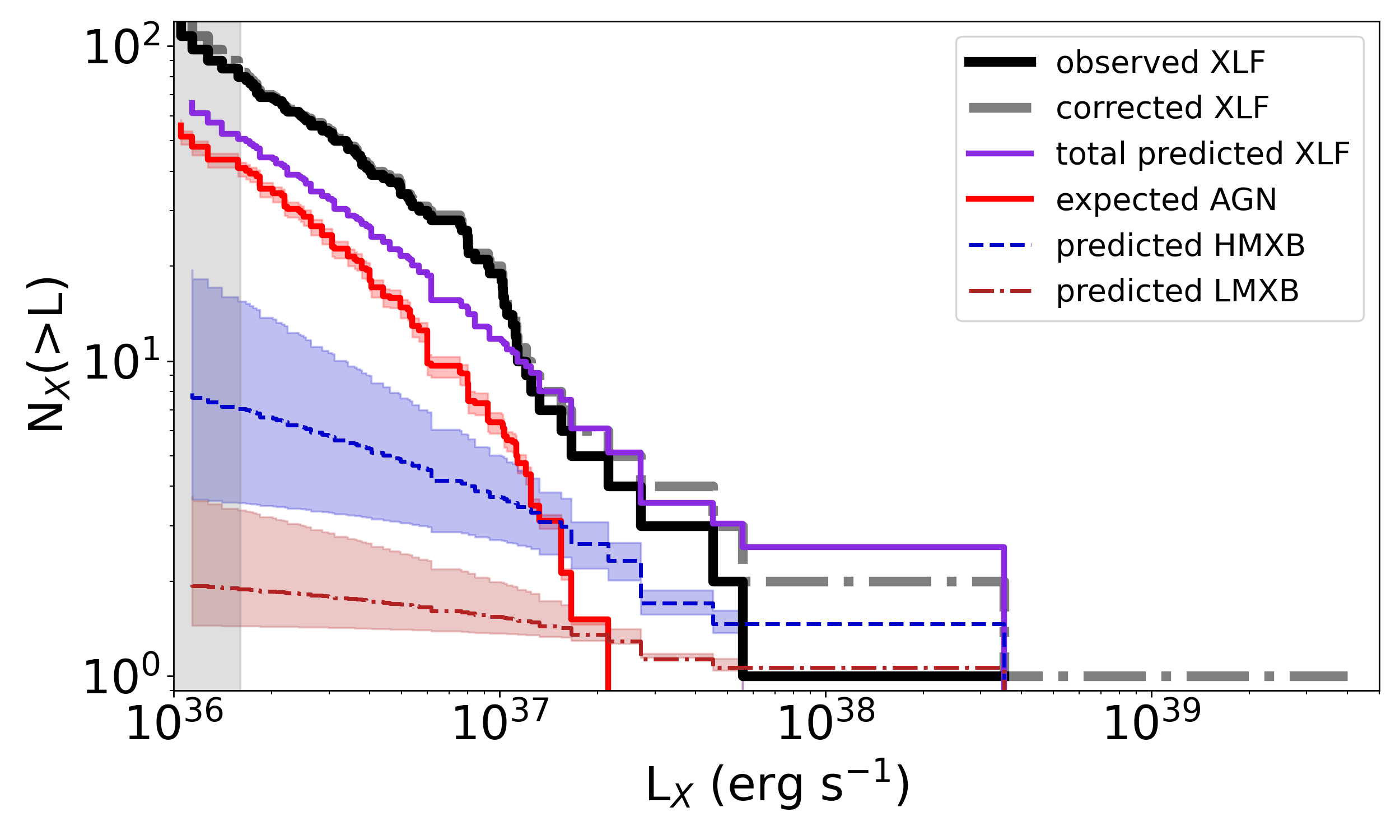} \\
    \includegraphics[width=1\linewidth]{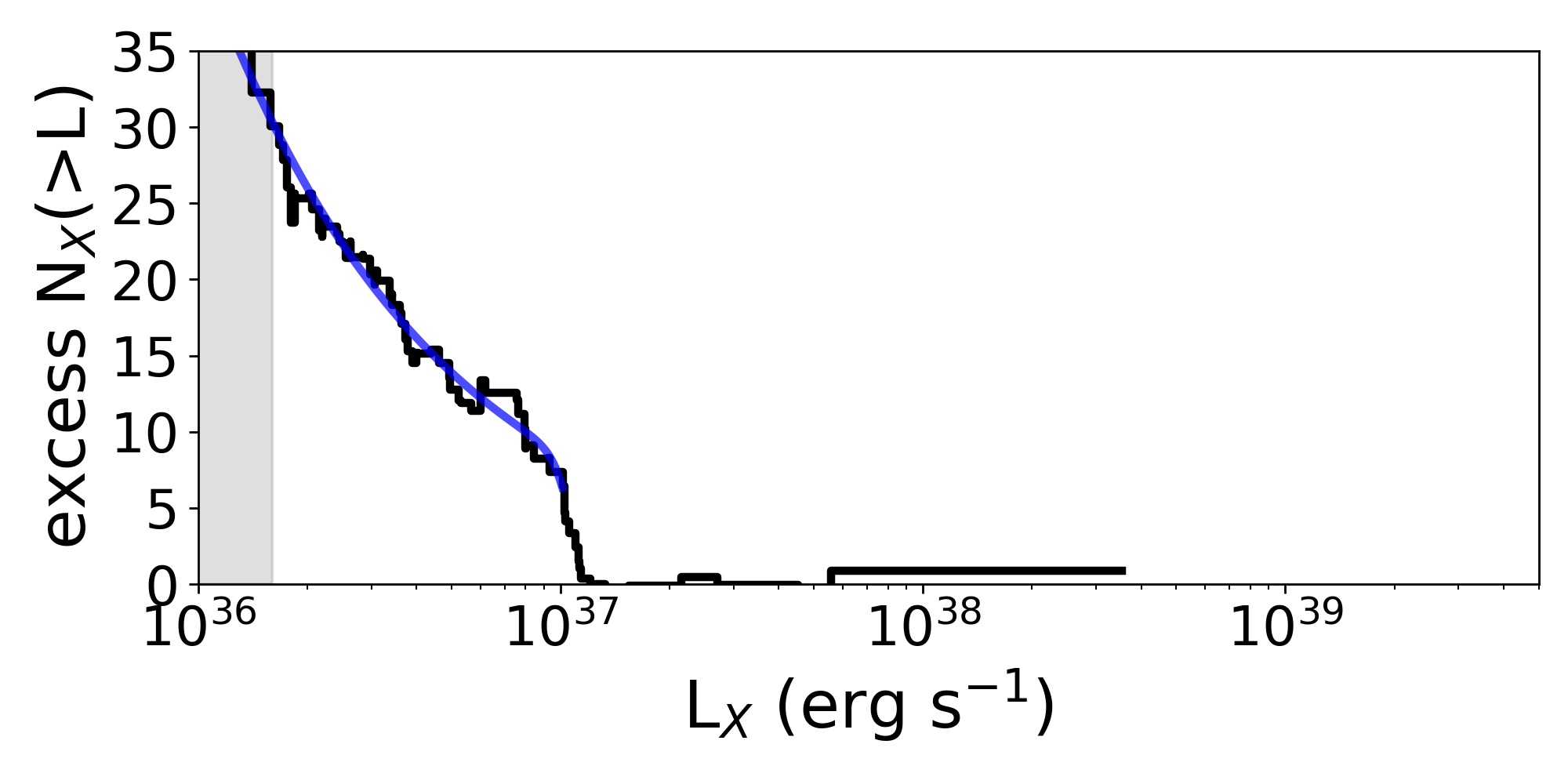} \\
    \caption{{\it Top}: The spatially-integrated XLF of NGC~300. The solid black line shows the observed XLF when a fiducial spectral model is used to directly convert count rates to luminosities, while the dot-dashed gray line shows the XLF after correcting for the ``true'' X-ray luminosity of NGC~300 ULX-1. The predicted contributions of AGN, HMXBs, and LMXBs are shown as solid red, dashed blue, and dot-dashed dark red lines, respectively (the shaded region indicates uncertainties), and the sum of these three components is shown in purple. {\it Bottom}: The ``excess'' XLF (the difference between the gray and purple distributions in the top panel) with a broken power law model (blue) superimposed. Only the horizontal axis uses a logarithmic scale, which makes the power law model appear curved. The gray shaded region indicates the 90\% limiting luminosity cutoff.}
    \label{fig:predicted_XLFs}
\end{figure}

The excess number of observed X-ray sources is also apparent in the XLF, as shown in Figure~\ref{fig:predicted_XLFs}. The unabsorbed luminosity of an X-ray source is dependent on the spectral model assumed; oftentimes, a fiducial spectral model appropriate for both XRBs and AGN is assumed (usually an absorbed power law) and used to convert count rates to unabsorbed luminosities. We first use WebPIMMS\footnote{\url{https://heasarc.gsfc.nasa.gov/cgi-bin/Tools/w3pimms/w3pimms.pl}} to implement this approach, assuming a power law model with a photon index $\Gamma=1.7$ that is subject to the Galactic absorbing column \citep[$N_{\rm H}=10^{21}$ cm$^{-2}$ towards NGC~300;][]{HI4PICollaboration+16} to convert the measured count rates to fluxes (and then into luminosities using the distance to NGC~300). The cumulative XLF can then be computed using

\begin{equation}
    N(>S) = \sum_{i,S_i>S} \frac{1}{A(S_i)},
\end{equation}

\noindent where $A(S)$ is the area over which a source with flux $S$ could be detected as measured from the sensitivity curves in Section~\ref{sec:sensitivity}. The resulting cumulative XLF is shown in black in Figure~\ref{fig:predicted_XLFs}. There is one striking example of a source in NGC~300 where the assumption of a fiducial spectral model fails to yield a realistic luminosity approximation: NGC~300 ULX-1. Detailed observations of this source by \xmm\ and {\em NuSTAR} indicate a relatively constant luminosity of $\sim4\times10^{39}$ \lum, with the modest count rates observed by \chandra\ likely the result of significant local obscuration by a precessing accretion disk \citep[see, e.g.][and references therein]{Binder+20}. We thus replace the fiducial model luminosity for NGC~300 ULX-1 with the ``true'' luminosity of the source, and recalculate a ``corrected'' XLF (dot-dashed gray line in Figure~\ref{fig:predicted_XLFs}). We additionally show the ChaMP AGN log$N$-log$S$ distribution \citep[red][]{Kim+04} and the synthetic HMXB (dashed blue) and LMXB (dot-dashed dark red) XLFs described in the previous section. The total predicted XLF (AGN + HMXB + LMXB) and the corresponding uncertainties are shown in purple. When we replace the incorrect inferred \Lx\ of NGC~300 ULX-1 (from the fiducial model) with the more accurate \Lx\ we find that galaxy-integrated $L_{\rm X, int}$ in 
Table~\ref{tab:Xray_sources} becomes ($4.43\pm0.04$)$\times10^{39}$ \lum. The value of $L_{\rm X, int}$ becomes (4.36$\pm$0.03) $\times10^{39}$ \lum\ in the radial bin containing NGC~300 ULX-1 ($R_d<r \leq 2R_d$).

There is excellent agreement between the ``corrected'' observed XLF and the predicted XLF above $\sim10^{37}$ \lum. Below this value, we observe $\sim$20 more X-ray sources than predicted based on the survey area, stellar mass, and SFR. If we attempt to describe the low-flux end of the XLF by increasing the AGN contribution, the assumed survey area must be increased by $\sim$60\%, which is well in excess of any area corrections derived from the uncertainties in $R_d$. Furthermore, increasing the AGN contribution leads to a clear excess of sources at $\sim(1-4)\times10^{37}$ \lum, where our survey is expected to be highly complete. 

We subtract the observed XLF from the predicted XLF and show the resulting ``excess'' XLF in the bottom panel of Figure~\ref{fig:predicted_XLFs}. Both a broken power law model and an exponential cutoff model describe the shape of this excess XLF well; in either case, the faint-end slope is $\sim$0.7 with a break at $\sim10^{37}$ \lum. In the broken power law model, the bright-end slope is very steep ($\gg$5) but largely unconstrained. There are two plausible explanations for the observed excess: (1) missing AGN, as the faint-end slope is broadly consistent with the faint end of the AGN log$N$-log$S$ distribution, and (2) variable sources, as the broken power law shape of the excess XLF matches the predicted shape of the XLF of simulated variable XRBs \citep{Binder+17}. The XRB scaling relations from Equation~\ref{eq:scaling_relations} were derived from on-average brighter XRB populations (largely above $10^{37}$ \lum) than those observed in the merged NGC~300 image and, as shown in Figure~\ref{fig:predicted_XLFs} (top panel) the uncertainties in the HMXB XLF at luminosities below $10^{37}$ \lum\ are significant. If we assume that all of the excess sources below $10^{37}$ \lum\ are XRBs, their total \Lx\ would be $\sim2.5\times10^{37}$ \lum, which is comparable to the uncertainties on the predicted $\mathcal{L}_X$ values from the XRB scaling relations (see Table~\ref{tab:Xray_sources}). Although the HMXB XLF (and corresponding \Lx/SFR scaling relation) adequately predicts the integrated \Lx\ of NGC~300, it does not predict the total number of objects in the galaxy below $\sim$10$^{37}$ \lum.

\section{Discussion}\label{sec:discussion}
Numerous works have investigated the relationship between the number and collective luminosity of HMXBs and host galaxy SFR and metallicity. However, the majority of these studies have utilized galaxy-integrated measurements of SFR and $Z$, and have largely targeted vigorously star-forming nearby galaxies to ensure a well-populated high-luminosity HMXB XLF. We now compare the results of our subgalactic analysis of NGC~300 to two key prior results: the $N_{\rm HMXB}$-SFR relationship \citep{Mineo+12} and the \Lx-SFR-$Z$ relationship. We seek to emphasize (1) that SFHs of more localized regions within a galaxy can vary, for example as a result of inside-out disk growth, which leads to measurable differences in the XRB populations that form; (2) that metallicity gradients provide a means to test metallicity dependence of observable XRB properties within a single galaxy; and (3) that HMXB-SFR scaling relations may not be reliable tools for predicting the XRB population below $\sim$10$^{37}$ \lum.

There is evidence that HMXBs, in particular, are related to even more local conditions than considered here: for example, \citet{Williams+13} used CMD fitting of the \hst-resolved stellar populations in the very local ($\sim$50 pc)  vicinity of HMXB candidates in NGC~300 and found that the SFRs of HMXB-hosting regions was a factor of $\sim$2.5 higher than that of randomly selected regions roughly $\sim$40-60 Myr ago. HMXB formation, however, is an intrinsically rare and stochastic process -- not all highly localized regions that experienced a similar increase in SFR $\sim$50 Myr produce an HMXB -- so additional assumptions about, e.g., the HMXB formation efficiency, are required to compared these results to the galaxy-integrated literature. We additionally examine whether highly localized ($\sim$100 pc) environments around XRB candidates exhibit significant differences in SFH or metallicity than the sub-galactic annular regions discussed above.

\subsection{The $N_{\rm HMXB}$-SFR Relationship}
Since our observations are not able to definitively separate HMXBs from LMXBs for a majority of sources in NGC~300, we assume that the {\it relative} number and luminosities of HMXBs to LMXBs follow the predictions of the XRB scaling relations for each subgalactic bin to estimate the number of HMXBs (and the integrated \Lx\ of these HMXBs) in each radial bin. In other words,

\begin{equation}\label{eq:N_HMXB}
    N_{\rm HMXB} = N_{\rm int}\times \left(\frac{\mathcal{N}_{\rm HMXB}}{\mathcal{N}_{\rm HMXB} + \mathcal{N}_{\rm LMXB}} \right)
\end{equation}

\noindent and

\begin{equation}
    L_{\rm HMXB} = L_{\rm int}\times \left(\frac{\mathcal{L}_{\rm HMXB}}{\mathcal{L}_{\rm HMXB} + \mathcal{L}_{\rm LMXB}} \right).
\end{equation}

\begin{figure}
    \centering
    \includegraphics[width=1\linewidth]{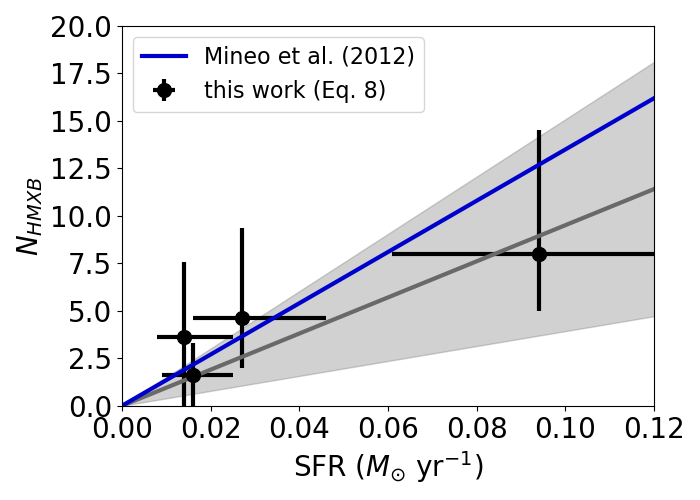}
    \caption{The number of HMXBs above $L_{\rm X,90}$ (given by Equation~\ref{eq:N_HMXB}) as a function of SFR in NGC~300. Black points show our data, the best-fit linear relationship (and the 3$\sigma$ uncertainty) is shown in gray. The solid blue line shows the relationship derived by \citet{Mineo+12}.}
    \label{fig:Nx_SFR_Mineo}
\end{figure}

\citet{Mineo+12} observed a linear relationship between the number of HMXBs observed in a galaxy and the host galaxy SFR. In that work, a single power law XLF was fit to the bright ($>10^{37}$ \lum) HMXB population and used to predict the total number of HMXBs above $10^{35}$ \lum. To compare with those results, in Figure~\ref{fig:Nx_SFR_Mineo} we plot the predicted number of HMXBs above $L_{X,90}$ as a function of SFR for the four subgalactic regions within NGC~300. A simple linear regression fit yields

\begin{equation}
    N_{\rm HMXB}(>10^{36}~\text{erg s}^{-1}) = \left(95\pm19 \right) \left(\frac{\text{SFR}}{M_{\odot} \text{ yr}^{-1}}\right).
\end{equation}

\noindent The slope of this relationship is in good agreement (within $\sim2\sigma$) with \citet[][slope of 135]{Mineo+12}, which is remarkable given the very different populations of sources considered and different assumptions invoked in the two studies. Assuming the predicted number of HMXBs, $\mathcal{N}_{\rm HMXB}$, derived from the \Lx/SFR scaling relation yields a significantly shallower slope: 43$\pm$5. We can additionally estimate the HMXB production rate by dividing the expected number of HMXBs in each subgalactic region by the total stellar mass formed in the last 100 Myr within each region. Across all four subgalactic regions within NGC~300, we estimate an HMXB production rate of $\sim$(1-2)$\times10^{-6}$ \Msun$^{-1}$. These values are lower than observed in M33 \citep{Lazzarini+23}, but are comparable to the production rates of HMXBs in M31 $\lesssim$50 Myr ago \citep{Lazzarini+21}. The agreement between expected number of HMXBs and the HMXB production rate to other dedicated HMXB studies suggests that the extra X-ray sources observed below $10^{37}$ \lum\ are dominated by HMXBs, and not AGN.

\begin{figure}
    \centering
    \includegraphics[width=1\linewidth]{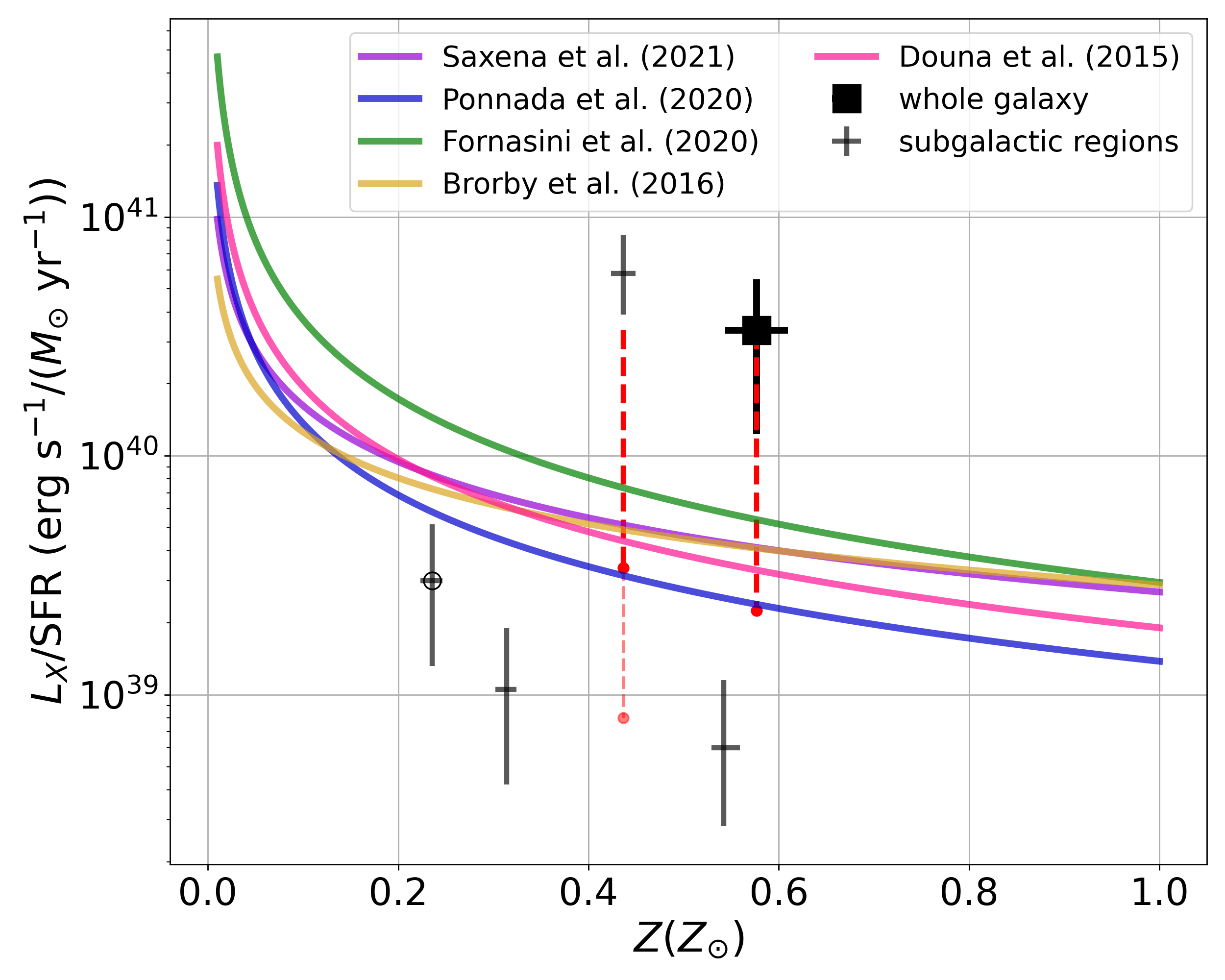}
    \caption{The ratio of \Lx/SFR as a function of metallicity for the four subgalactic regions in NGC~300 (gray) and the whole galaxy within $4R_d$ (black) compared to recent \Lx-SFR-$Z$ relationships from the literature. Uncertainties in the literature relationships are typically on the order of $\sim$0.5 dex. The point at $Z\approx0.43Z_{\odot}$ indicates the subgalactic region containing the two brightest (and youngest) sources in NGC~300: NGC~300 X-1 and NGC~300 ULX-1. The vertical bright, dashed red lines show the effect of removing NGC~300 ULX-1 from our analysis, while the faint, dashed red line shows the effect of additionally removing NGC~300 X-1. The only subgalactic region with a constant SFR between 0-30 Myr and 30-100 Myr is indicated with an open circle.}
    \label{fig:LX-SFR-Z}
\end{figure}

\subsection{The \Lx-SFR-$Z$ Relationship}

In Figure~\ref{fig:LX-SFR-Z} we compare our subgalactic (and galaxy-integrated) observations of NGC~300 to several \Lx-SFR-$Z$ relationships from the literature (uncertainties in the literature relationships are typically on the order of $\sim$0.5 dex). To estimate the \Lx\ originating from HMXBs only, we take the observed $L_{\rm X, int}$ from Table~\ref{tab:Xray_sources} and multiply it by the percentage of the predicted $\mathcal{L}_{\rm X}$ originating in HMXBs from the XRB scaling relations. The predicted $\mathcal{L}_{\rm X}$ was determined from simulating XRB populations from within the stellar mass and SFRs for NGC~300, so the uncertainties on $\mathcal{L}_{\rm X}$ are inclusive of both measurement uncertainties and additional uncertainties that result from stochastic formation of very bright sources (Section~\ref{sec:XRBs}). These uncertainties are comparable to the stochastic uncertainties in \Lx/SFR reported by \citet{Lehmer+21}. We find that the galaxy-integrated \Lx/SFR ratio of NGC~300 agrees with predictions for the value of $Z$ we measure from our SED modeling if the corrected \Lx\ of NGC~300 ULX-1 is not used; using the higher \Lx\ for this one source puts the galaxy-integrated \Lx/SFR ratio $\sim1\sigma$ above the literature relationships. Two of the four subgalactic regions fall below the literature \Lx-SFR-$Z$ relationships, while the outermost region (at the lowest $Z$) is consistent (within $\sim1\sigma$) with the literature relationships. The only radial bin that lies above the literature results is the bin from $R_d<r\leq 2R_d$, which has the highest recent SFR and contains the two youngest and brightest HMXBs in NGC~300: ULX-1 and X-1. If we exclude NGC~300 ULX-1 from our analysis, the predicted $\mathcal{L}_{\rm X}$ for HMXBs from the XRB scaling relations in Equation~\ref{eq:scaling_relations} is consistent with the observed \Lx\ of NGC~300 X-1 and the \Lx/SFR ratio for this radial bin becomes consistent with literature \Lx-SFR-$Z$ relationships (bright red, dashed line in Figure~\ref{fig:LX-SFR-Z}).

Figure~\ref{fig:LX-SFR-Z} thus demonstrates the effect that stochastic formation of bright HMXBs in modestly star forming galaxies can have on the observed \Lx/SFR ratio. Many ULXs of comparable \Lx\ to NGC~300 ULX-1 are located within galaxies with ongoing and significantly higher SFRs (often by an order of magnitude or more) than that observed in NGC~300, which is past the peak of its SFH. When NGC~300 ULX-1 is excluded, the subgalactic region at $R_d<r\leq 2R_d$ ($Z\approx0.4Z_{\odot}$) is the only radial bin consistent with literature \Lx-SFR-$Z$ relationships (bright, dashed red line in Figure~\ref{fig:LX-SFR-Z}). The remaining \Lx\ from this region is dominated by the Wolf-Rayet + black hole binary NGC~300 X-1 \citep{Binder+21}, which is assumed to be extremely young based on the high masses of both the black hole and donor star. The outermost radial bin ($Z\approx0.25Z_{\odot}$) falls below the literature \Lx-SFR-$Z$ relationships only by $\sim1-2\sigma$. This is the only subgalactic region where the SFR increased (by $\sim$30\%) in the last 30 Myr compared to 30-100 Myr ago, and so likely contains a higher proportion of younger, more massive HMXBs compared to other regions of NGC~300.

These differences in the \Lx/SFR ratios of the subgalactic regions within NGC~300 may be hinting at an HMXB stellar age effect on the \Lx-SFR-$Z$ relationship, although with a limited number of data points it is difficult to draw any firm conclusions. That X-ray emission from XRB populations evolves as a function of time is not new \citep{Fragos+13a,Lehmer+19,Antoniou+19}, but this effect is difficult to directly measure on the short timescales over which HMXBs evolve ($\lesssim$100 Myr). The XLFs of LMXBs reveal that older galaxies tend to host a larger number of faint LMXBs ($\lesssim10^{37}$ \lum) compared to younger galaxies, and that frequency of bright- and ULX-class LMXBs is much higher in younger galaxies \citep{Zhang+12}. The duration of the HMXB phase for a single binary is expected to be short \citep[$\sim$10$^4$ yr,][]{Mineo+12}. If more massive HMXBs are systematically more luminous than lower-mass HMXBs, one would expect the the X-ray emission from an HMXB population to shift from younger/brighter systems to older/fainter systems over time, and in the absence of renewed star formation episodes the collective \Lx\ would diminish as brighter systems begin to ``turn off.'' The evolution of \Lx/\Mstar\ in star-forming galaxies over cosmic time has been examined both observationally \citep{Gilbertson+22} and theoretically \citep{Linden+10,Fragos+13b,Fragos+13a}, and this study of NGC~300 which approximately controls for stellar age on timescales $<$100 Myr provides another framework by which HMXB age evolution could be studied further.

Prior studies have systematically selected higher-SFR galaxies with ongoing star formation for analysis, which results in a proportionally larger population of young and luminous HMXBs (akin to NGC~300 ULX-1 and NGC~300 X-1) compared to the more ``typical,'' moderate-luminosity $\sim$40-60 Myr HMXBs that are readily detectable in nearby galaxies \citep[in NGC~300, but also the Magellanic Clouds, M31, M33, and NGC~2403;][]{Williams+13,Lazzarini+21,Lazzarini+23,Antoniou+16,Antoniou+10}. We note that if NGC~300 X-1 were to evolve out of its current X-ray bright phase, the \Lx/SFR ratio of its corresponding subgalactic region would follow the general downward trend towards higher metallicities seen in the other three radial bins (faint red, dashed line in Figure~\ref{fig:LX-SFR-Z}). While the stochastic formation of very bright and short-lived HMXBs is expected to be influenced by metallicity, the current understanding of the \Lx-SFR-$Z$ relationship and XRB scaling relations are likely biased towards this small subset of younger, brighter HMXBs. Whether increasing stellar ages result in an \Lx-SFR-$Z$ relationship with a different normalization or slope (or both) compared to the one derived from presumably younger HMXBs is uncertain based on the present study; a more rigorous test of the effect of stellar age on the HMXB \Lx-SFR-$Z$ relationship would require a comprehensive study of the fainter HMXB populations in modestly star-forming galaxies with spatially resolved SFHs and metallicity gradients. 

Evidence of variation and evolution in the \Lx-SFR-$Z$ relationship may be hidden within the significant scatter seen in previous studies. The literature \Lx-SFR-$Z$ relationships were derived using a variety of SFR indicators and stellar IMF assumptions, and some of the scatter in these relationships may originate from a heterogeneous mix of SFR tracers in addition to different assumptions about metallicity calibrations. Self-consistently modeling the modern SFRs, SFHs, and metallicities in subgalactic regions concurrently with the HMXBs population is required to firmly establish the extent to which stochastic and stellar age effects contribute to the \Lx-SFR-$Z$ relationship, and ultimately our understanding of massive binary star evolution.

\subsection{Comparison to XRB-Hosting Local Regions}
The short lifetimes of HMXBs ($\lesssim$100 Myr) and relatively low natal kick velocities they experience during initial compact object formation \citep{Giacobbo+20} prevent HMXBs from travelling more than $\sim$100-200 pc from their birthplaces. We thus explored whether the environments directly associated with XRB candidate sources showed any significant differences in terms of their SFHs or metallicities compared to the larger annular regions described above.

We identified XRB candidate X-ray sources by excluding highly absorbed sources (e.g., those with HR1 $\geq$ 0.4 which are likely background AGN) and extracted photometry from circular apertures centered on the X-ray sources with radii of 10.3$^{\prime\prime}$ ($\sim$100 pc). We verified that none of the individual circular apertures overlapped; thus the total area covered is 0.03 kpc$\times$ the number of XRB candidates in each annular region (from innermost to outermost region, we identify 4, 11, 7, and 10 XRB candidates). The SED resulting from the combined photometry from all XRB candidates within a given annular region is then modeled with \prospector\ in a manner identical to that described in Section~\ref{sec:SEDs}.

\begin{figure*}
    \centering
    \begin{tabular}{ll}
        \includegraphics[width=0.6\linewidth,trim=0cm 3.1cm 0cm 0cm]{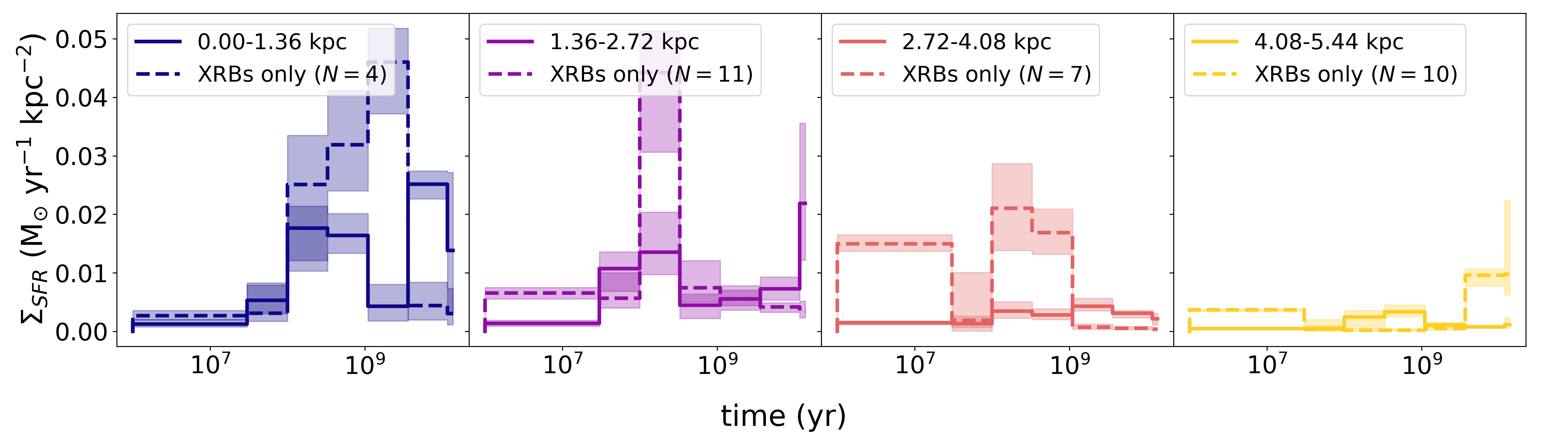} & 
        \multirow[c]{2}{*}[1.7cm]{\includegraphics[width=0.4\linewidth,trim=1cm 0cm 0cm 0cm]{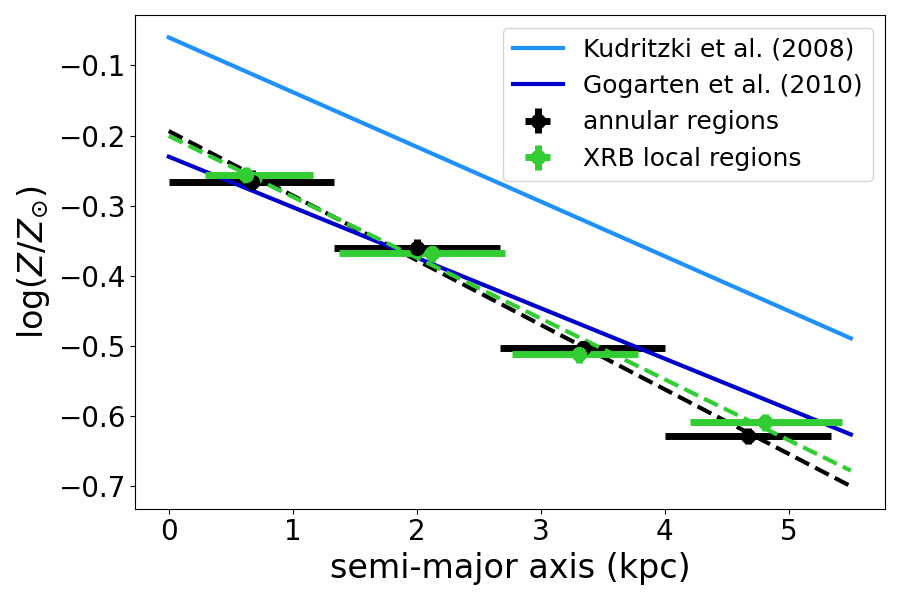}}    \\
        \includegraphics[width=0.6\linewidth]{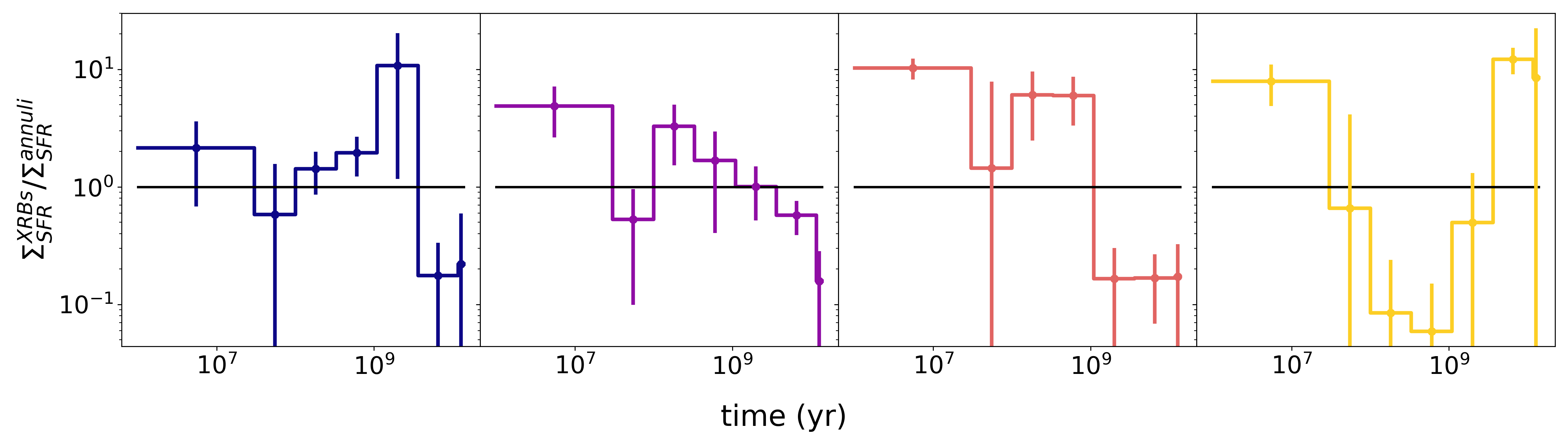} & \\
    \end{tabular}
    \caption{{\it Left}: The radially-resolved SFHs of the annular regions (solid lines) compared to the stacked localized regions around XRB candidates (dashed lines; the number of XRB candidates used to derive the SFHs is indicated in the legend). Colors are as in Figure~\ref{fig:SFH_comparison}. {\it Right}: The metallicity gradient derived from the stacked localized regions containing XRBs (green) compared to the annular regions (black), \citet[][light blue]{Kudritzki+08}, and \citet[][dark blue]{Gogarten+10}. }
    \label{fig:localized_regions}
\end{figure*}

The resulting SFHs and the metallicity gradients from both the XRB-hosting local regions and the larger annular regions are shown in Figure~\ref{fig:localized_regions}. To account for the significant differences in areas covered, we show the SFR surface density $\Sigma_{\rm SFR}$. The SFHs of the XRB-hosting local regions generally show an enhancement in $\Sigma_{\rm SFR}$ the most recent ($<$30 Myr) time bin by factors of $\sim$3-10 while the SFRs observed in the 30-100 Myr time bins are consistent (within the large uncertainties) with those found from the annular analysis presented. This leads to an overall increase in the inferred SFR within the last 100 Myr in the locally-derived SFR compared to the annuli-averaged SFR. The metallicity gradient derived from the XRB-hosting regions is nearly identical to the one derived from the larger annular regions. If the SFRs associated with XRBs is indeed higher than the annuli-averaged value, the \Lx/SFR ratios shown on Figure~\ref{fig:LX-SFR-Z} would be shifted down. The deviation between the NGC 300 subgalactic analysis and the literature \Lx-SFR-$Z$ relationships becomes more pronounced; the NGC~300 sources are underluminous for their metallicity and SFR. This is consistent with the finding in Section~\ref{sec:obs_vs_pred} (see also Figure~\ref{fig:predicted_XLFs}), where we find evidence for an excess of possibly variable, low-luminosity XRB population in excess of the predictions from HMXB and LMXB scaling relations.

\section{Summary}\label{sec:conclusion}
We have conducted a panchromatic analysis of the nearby galaxy NGC~300 on subgalactic scales to illustrate how spatially-resolved studies of nearby galaxies can provide useful insights into the XRB-host galaxy connection. The 3.6$\mu$m scale length ($R_d$) is used to divide the galaxy into four equal-width annular regions, and we perform aperture photometry to measure the UV-optical-IR flux densities within each subgalactic region. We then construct SEDs and use \prospector\ to measure the SFHs and present-day stellar masses, metallicities, and recent ($<$100 Myr) SFRs across the star-forming disk. The current galaxy-integrated stellar mass is \Mstar=$(2.15^{+0.26}_{-0.14})\times10^9$ \Msun\ and the current SFR is $\sim$0.18$\pm$0.07 \Msun\ yr$^{-1}$, in agreement with previous measurements. Our measured metallicity gradient and radially-resolved SFH are also consistent with other gradients reported in the literature.

We used four archival \chandra\ observations to identify 70 X-ray sources within $4R_d$ that were detected at high significance. We used the ChaMP AGN log$N$-log$S$ distribution \citep{Kim+04} and the XRB scaling relations of \citet{Lehmer+20,Lehmer+21} to compare the observed X-ray point source population to the predicted X-ray point source population. On a galaxy-integrated scale, the radial source distribution and XLF show a clear excess in the number of observed X-ray sources at luminosities $<10^{37}$ \lum. The X-ray source excess is consistent with a faint and variable XRB population that is not well described by the assumed XRB scaling relations, with the possibility of additional AGN contamination. Our subgalactic analysis reveals the linear relationship between the number of HMXBs formed and SFR that is consistent with prior expectations that derived from a large, galaxy-integrated sample of more vigorously star forming galaxies. 

Most intriguingly, we find that the \Lx/SFR ratios as a function of $Z$ across the four subgalactic regions are only consistent with \Lx-SFR-$Z$ relationships from the literature for the bins that are likely to host the youngest HMXBs. These results hint at a possible trend in the \Lx-SFR-$Z$ relationship with stellar age, as lower-mass HMXBs begin to dominate the total \Lx\ of a galaxy after the star formation episode that formed those systems ends. We derive SFHs for the local ($\sim$100 pc) regions immediately surrounding XRB candidate sources and find that their recent ($<$30 Myr) SFR surface densities are generally higher than those of the integrated annular regions, which decreases the \Lx/SFR ratio and makes the underluminous nature of the NGC~300 XRB population even more pronounced. Expanding this analysis to a larger sample of nearby galaxies that are just past their peak of star formation, and where the fainter X-ray source population can be studied on subgalactic scales (and the spatially-resolved SFH can be directly modeled), is needed to further investigate possible trends in the \Lx/SFR ratio with both metallicity and stellar age.

\begin{acknowledgments}
B.A.B acknowledges support from the National Science Foundation Launching Early-Career Academic Pathways in the Mathematical and Physical Sciences (LEAPS-MPS) award \#2213230. This paper employs a list of \chandra\ datasets, obtained by the \chandra\ X-ray Observatory, contained in~\dataset[DOI: 10.25574/cdc.204]{https://doi.org/10.25574/cdc.204}, and software provided by the \chandra\ X-ray Center (CXC) in the application package CIAO. This research made use of Astropy, a community-developed core Python package for Astronomy \citep{astropy:2013, astropy:2018, astropy22}.
\end{acknowledgments}

\facilities{CXO, Swift (UVOT), GALEX, 2MASS, Spitzer (IRAC and MIPS), WISE}

\software{astropy \citep{astropy:2013,astropy:2018,astropy22}, CIAO \citep{Fruscione+06}
          }

\bibliography{references}{}
\bibliographystyle{aasjournal}

\end{document}